\documentclass[letterpaper,11pt]{article}
\usepackage{jheppub} 
\pdfoutput=1
\usepackage{amsmath}
\usepackage{xcolor}
\usepackage{youngtab}
\usepackage{subcaption}

\usepackage{graphicx}
\usepackage{verbatim}
\usepackage{framed}
\definecolor{shadecolor}{rgb}{0.90,0.90,0.90}
\def\beq{\begin{eqnarray}}\def\eeq{\end{eqnarray}}
\def\be{\begin{equation}}\def\ee{\end{equation}}
\def\g{\gamma}

\def\s{\sigma}
\def\m{\mu}
\def\n{\nu}
\def\a{\alpha}

\def\b{\beta}
\def\d{\delta}

\def\D{\Delta}

\def\3s{{s \choose 3}}
\def\4s{{s \choose 4}}
\def\5s{{s \choose 5}}
\def\6s{{s \choose 6}}

\def\12{\dfrac{1}{2}}

\def\2{\ell_2}

\def\be{\begin{equation}}
\def\ee{\end{equation}}
\def\bea{\begin{eqnarray}}
\def\eea{\end{eqnarray}}
\def\ba{\begin{array}}
	\def\ea{\end{array}}
\def\bec{\begin{center}}
	\def\ec{\end{center}}

\def\g{\gamma}

\def\s{\sigma}
\def\m{\mu}
\def\n{\nu}
\def\a{\alpha}

\def\b{\beta}
\def\d{\delta}

\def\D{\Delta}

\def\3s{{s \choose 3}}
\def\4s{{s \choose 4}}
\def\5s{{s \choose 5}}
\def\6s{{s \choose 6}}

\def\12{\dfrac{1}{2}}

\def\2{\ell_2}

\def\be{\begin{equation}}
\def\ee{\end{equation}}
\def\bea{\begin{eqnarray}}
\def\eea{\end{eqnarray}}
\def\ba{\begin{array}}
	\def\ea{\end{array}}
\def\bec{\begin{center}}
	\def\ec{\end{center}}

\def\a{\alpha} 
\def\b{\beta}  
\def\g{\gamma} 

\def\d{\delta} 
\def\D{\Delta}

\def\m{\mu}
\def\n{\nu}

\def\s{\sigma}

\newcommand{\Z}{\mathbb{Z}}

\def\tilde{\widetilde}

\def\bar{\overline}

\def\CM{{\mathcal M}}

\def\CS{{\mathcal S}}

\def\CV{{\mathcal V}}

\def\cf{\mathcal{F}}

\def\ts{{\mathtt s}}
\def\tv{{\mathtt v}}

\newcommand{\polo}{\epsilon^\perp}

\newcommand{\syng}[1]{\scalebox{0.2}{\yng(#1)}}

\definecolor{nicered}{rgb}{0.7,0.1,0.1}
\definecolor{nicegreen}{rgb}{0.1,0.5,0.1}

\def\denom{{\mathtt D}}
\definecolor{mGreen}{rgb}{0,0.6,0}
\definecolor{mgray}{rgb}{0.6,0.6,0.6}
\definecolor{mpurple}{rgb}{0.58,0,0.82}
\definecolor{backgroundColour}{rgb}{0.95,0.95,0.92}

\definecolor{mred}{rgb}{0.5,0.0,0.0}
\definecolor{mgreen}{rgb}{0.0,0.4,0.0}
\definecolor{mblue}{rgb}{0.0,0.0,0.6}
\definecolor{myellow}{rgb}{0.4,0.4,0.0}
\definecolor{mpink}{rgb}{0.4,0.0,0.4}
\definecolor{mcyan}{rgb}{0.0,0.4,0.4}
\definecolor{mblack}{rgb}{0.0,0.0,0.0}

\def\ve{{\varepsilon}}
\def\s{{\sigma}}
\def\g{{\gamma}}

\def\a{{\alpha}}
\def\b{{\beta}}

\def\d{{\delta}}

\def\CM{{\mathcal M}}

\def\CS{{\mathcal S}}

\def\CV{{\mathcal V}}

\newcommand{\bd}[1]{\begin{fmffile}{#1}\begin{fmfgraph*}}
		\newcommand{\ed}{\end{fmfgraph*}\end{fmffile}}

\def\0{{(0)}}
\def\1{{(1)}}
\def\2{{(2)}}
\def\3{{(3)}}
\def\4{{(4)}}

\def\+{{(+)}}
\def\-{{(-)}}

\def\be{\begin{equation}}
\def\ee{\end{equation}}
\def\beq{\be\begin{array}{c}}
	\def\eeq{\end{array}\ee}

\preprint{TIFR/TH/20-17}

\title{Classification of four-point local gluon S-matrices}

\author{Subham Dutta Chowdhury, Abhijit Gadde}
\affiliation{Department of Theoretical Physics,\\ Tata Institute of Fundamental Research, Mumbai 400005, India
}

\emailAdd{subham@theory.tifr.res.in, abhijit@theory.tifr.res.in}

\abstract{
	In this paper, we classify four-point local gluon S-matrices in arbitrary dimensions. This is along the same lines as \cite{Chowdhury:2019kaq} where four-point local photon S-matrices and graviton S-matrices were classified. We do the classification explicitly for gauge groups $SO(N)$ and $SU(N)$ for all $N$ but our method is easily generalizable to other Lie groups. The construction involves combining not-necessarily-permutation-symmetric four-point S-matrices of photons and those of adjoint scalars into permutation symmetric four-point gluon S-matrix. We explicitly list both the components of the construction, i.e permutation symmetric as well as non-symmetric four point S-matrices, for both the photons as well as the adjoint scalars for arbitrary dimensions and for gauge groups $SO(N)$ and $SU(N)$ for all $N$. In this paper, we explicitly list the local Lagrangians that generate the local gluon S-matrices for $D\geq 9$ and present the relevant counting for lower dimensions. Local Lagrangians for gluon S-matrices in lower dimensions can be written down following the same method. We also express the Yang-Mills four gluon S-matrix with gluon exchange in terms of our basis structures.   
}

\begin{document}

\maketitle
\flushbottom

\section{Generalities}


Consider a Yang-Mills theory in $D$ dimensions with gauge group $G$. In this paper, we will take $G$ to be either $SO(N)$ or $SU(N)$, although generalization to any Lie group is straightforward. We are interested in classifying the four-point gluon S-matrix. In this section, we will review generalities of the $n$-point gluon S-matrix.
The $n$-point gluon scattering amplitude is a function $\CS(\epsilon_{\mu}^{(i),a},p_{\mu}^{(i)})$ of  polarizations $\epsilon_{\mu}^{(i),a}$ and momenta $p_{\mu}^{(i)}$.\footnote{The discussion of gluon scattering amplitudes in $D=4$ have a convenient description in terms of the so called spinor-helicity variables. See \cite{Elvang:2013cua} for a review of and an extensive list of references on gluon scattering in $D=4$.
As we are interested in classifying gluon scattering in arbitrary number of dimensions, it serves use well to stick with the use of more conventional variables: polarizations and momenta.}
Here $\mu$ and $a$ is the Lorentz and $G$-adjoint color index respectively. The superscript $(i)$ labels each gluon and goes from $1,\ldots, n$. The S-matrix is homogeneous with degree one with respect to each of the polarizations $\epsilon_{\mu}^{(i),a}$. It is defined for when external particles are massless and momentum is conserved. 
\be
(p^{(i)})^2=0,\qquad \qquad \sum_i\,  p_\mu^{(i)}=0.
\ee

The S-matrix is also invariant under certain transformations thanks to the gauge invariance of the action. 
The gauge invariance of the action under constant gauge transformation implies that the S-matrix is a $G$-singlet under the simultaneous gauge group transformations, 
\be
\epsilon_{\mu}^{(i),a} \to R^a_{\, b}\, \epsilon_{\mu}^{(i),b}.
\ee
The invariance of the action under non-constant gauge transformations implies the invariance of the S-matrix under under individual 
\be
\epsilon_{\mu}^{(i),a} \to \epsilon_{\mu}^{(i),a} +p_\mu^{(i)} \zeta^{(i),a}.
\ee
where $\zeta^{(i),a}$s are independent infinitesimal gauge transformations. It is useful to impose this invariance by thinking of the adjoint valued polarization vector as a product $\epsilon_{\mu}^{a}=\epsilon_{\mu} \otimes \tau^a$. The the S-matrix is invariant under the transformations of the separate variables,
\be
\tau^{(i),a}\to R^a_{\, b} \, \tau^{(i),b},\qquad\qquad  \epsilon_{\mu}^{(i)} \to \epsilon_{\mu}^{(i)} +p_\mu^{(i)} \zeta^{(i)}.
\ee
Then it is convenient to think of the gluon S-matrix as the sum of products,
\be\label{tensor}
\CS(\epsilon_{\mu}^{(i),a},p_{\mu}^{(i)}) =  \, \CS_{\rm photon} (\epsilon_{\mu}^{(i),a},p_{\mu}^{(i)})\, \CS_{\rm scalar}(\tau^{(i),a})+\ldots 
\ee
We recognize each term in the sum as the product of photon S-matrix and the S-matrix of adjoint scalar particles. In other words, $\CS_{\rm photon} \otimes \CS_{\rm scalar}$ gives a ``basis" for $\CS_{\rm gluon}$. This is made precise in the rest of the section. A general gluon S-matrix is then a sum of such basis elements. This is the sum appearing in \eqref{tensor}.   The scalar S-matrix is evaluated at zero momentum so it is really just a color structure i.e. a $G$-singlet.  
In addition to being invariant under Lorentz transformations and gauge transformations, the S-matrix of $n$ identical particles is also invariant the permutation symmetry $S_n$. 
In \ref{perm-inv} we will discuss imposition of the permutation symmetry on the tensor product structure for the case of $n=4$.

The Lorentz invariant functions of momenta are conveniently parametrized as functions of the so called Mandelstam variables. For the case of four massless particles these are,
\be
s=-(p_1+p_2)^2,\qquad t=-(p_1+p_3)^2,\qquad u=-(p_1+p_4)^2.
\ee
Momentum conservation implies $s+t+u=0$. For $n$ massless particles, the Mandelstam variables are
\be
s_{ij}=-(p_i+p_j)^2.
\ee
Note that $s_{ii}=0, s_{ij}=s_{ji}$ and $\sum_j \, s_{ij}=0$. This makes them $n(n-3)/2$ in number. Moreover, they transform under the following representation under the permutation group $S_n$. 
\be
\young(\,\,\,{\,\ldots\,\,\,\,}\,,\,\,)
\ee
At this point, we observe a qualitative difference between $n=4$ and $n>4$. For $n>4$, the above is a faithful representation of $S_n$ but for $n=4$, it has the kernel ${\mathbb Z}_2\times {\mathbb Z}_2$. This ${\mathbb Z}_2\times {\mathbb Z}_2$ is generated by double transpositions. Defining $P_{ij}$ to be an element of $S_4$ that transposes particles $i$ and $j$, the $\Z_2\times \Z_2$ consists of
\be\label{v4list}
\{1, \, P_{12}P_{34},\, P_{13}P_{24},\, P_{14}P_{23}\}.
\ee
The elements $P_{12}P_{34},\, P_{13}P_{24}$ can be taken to be the generators of the two $\Z_2$'s. The last entry $P_{14}P_{23}$ is the product of these generators.
The quotient $S_4/(\Z_2\times \Z_2)=S_3$. The $S_3$ permutes particles $2,3$ and $4$ and keeps  particle $1$ fixed. Mandelstam variables do form a faithful representation of the quotient group $S_3$. This representation is ${\bf 2_M}$.  In terms of the Young diagram,
\be
{\bf 2_M}=\yng(2,1).
\ee
See appendix \ref{s3} for review of the basics of  $S_3$ representation theory.
 In this paper, we will be interested only in the case $n=4$.

\subsection{Permutation symmetry: Module of quasi-invariants}\label{perm-inv}
The local  scattering amplitude of four identical gluons is invariant under the permutation $S_4$ of the external particles. As the Mandelstam variables $(s,t)$ are invariant under the normal subgroup $\Z_2\times \Z_2$ of $S_4$, it turns out to be invariant to impose the $S_4$ invariance in two steps. First impose invariance under $\Z_2\times \Z_2$ and then under the ``remnant" permutation group $S_3=S_4/(\Z_2\times \Z_2)$. We call the S-matrices that are invariant only under $\Z_2\times \Z_2$ subgroup, quasi-invariant. The advantage of imposing $S_4$ invariance in two steps is that the space of S-matrices obtained after the first step, viz. the space of quasi-invariant S-matrices is a finite dimensional vector space over the field of functions of $(s,t)$. This is because Mandelstam variables are quasi-invariant. This space admits an explicit characterization in terms of its basis vectors. Once we characterize this space, the second step is relatively straightforward.

In this paper, we are interested in the spacial class of S-matrices that we call local S-matrices. These S-matrices are polynomials in momenta as opposed to being general functions of momenta. As a result, the space of quasi-invariant local S-matrices is not a vector space but rather  a module over the ring of polynomials of $(s,t)$. A ring has a richer structure compared to a vector space. It can be described in terms of a generators $g_i$ such that all the elements of the module can be written as a linear combination $\sum_i r_i \cdot g_i$ where $r_i$ are elements of the associated ring. If all the elements of the ring are represented as such a combination uniquely then the module is a free module and  set of generators $g_i$ is said to generate it freely. If the module is a free module then it is characterized by its generators $g_i$, if it is not a free module then it is characterized by relations $\sum_i r_i \cdot g_i=0$ along with the generators $g_i$.\footnote{For complicated modules, one may further need to characterize relations of relations and so on. This description of a module is called the free resolution. In our case, all the non-free modules will be characterized by only by relations on generators. The relations themselves will be free i.e. they won't have any further relations.} Classification of gluon S-matrices is tantamount to characterizing the module of quasi-invariant S-matrices through generators and relations. The module of quasi-invariant S-matrix enjoys the action of $S_3$. We will always describe the generators and the relations by decomposing their space (thought of as a vector space over $\mathbb C$) them into irreducible representations of $S_3$.

\subsubsection{Partition function}\label{partition}
In this paper, it will be convenient for us to enumerate the generators and relations (if present) of all the quasi-invariant modules while keeping track of their $S_3$ representation and derivative order. This information can alternatively be encoded in a partition function over local physical i.e. $S_4$ invariant S-matrices, 
\be
Z_{\rm S-matrix}={\rm Tr} \,x^{\partial}.
\ee
where $\partial$ is the overall momentum homogeneity. 
As argued in \cite{Chowdhury:2019kaq}, local physical S-matrices are in one-to-one correspondence with the equivalence classes of quartic Lagrangians. The Lagrangians are said to be equivalent if their difference either vanishes on-shell or is a total derivative. The partition function over such equivalence classes of Lagrangians can be computed efficiently using plethystic integration techniques. For adjoint scalars and gluons, we will do so in section \ref{plethystic}. The partition function thus obtained serves as a check over explicitly construction of generators and relations. In this section, we establish a dictionary between the $S_3$ representation ${\bf R}$ and derivative order $\partial$ of generators (and relations) of the quasi-invariant module and  the partition function over local physical S-matrices. The $S_4$ invariant projection to get the physical S-matrix can be thought of simply as the $S_3$ invariant projection because the module elements are defined to be $\Z_2\times \Z_2$ invariant.

Consider a generator $|e_{\bf R}\rangle$ of derivative order $\partial$ transforming in an irreducible representation ${\bf R}$ of $S_3$. The $S_3$ invariant local S-matrix is obtained by taking its ``dot product" with a polynomial of $(s,t)$ that also transforms exactly in representation ${\bf R}$. Hence the partition function over the $S_3$ projection of the submodule of $|e_{\bf R}\rangle$ is,
\be
Z_{e_{\bf R}}(x)=Z_{\bf R}(x) x^{\partial}, 
\ee
where $Z_{\bf R}(x)$ is the partition function over polynomials of $(s,t)$ transforming in representation ${\bf R}$. There are only three irreducible representations of $S_3$. The partition functions $Z_{\bf R}(x)$ for all of them are given in \cite{Chowdhury:2019kaq}. We reproduce them below.
\be\label{Z-S3}
Z_{\bf S}=\frac{1}{(1-x^4)(1-x^6)},\quad Z_{\bf M}=\frac{x^2+x^4}{(1-x^4)(1-x^6)},\quad Z_{\bf A}=\frac{x^6}{(1-x^4)(1-x^6)}.
\ee
These expressions have a simple interpretation. The partition function $Z_{\bf S}$ is the partition function over totally symmetric polynomials of $(s,t)$. They are generated by $s^2+t^2+u^2$ and $stu$. The S-matrices from generators transforming in ${\bf 2_M}$ are made by first constructing two generators in ${\bf 1_S}$ and multiplying them by symmetric polynomials of $(s,t)$. Letting the generators of  ${\bf 2_M}$ be $|e^{(i)}_{\bf M}\rangle, i=1,2,3$  which transform as naturally as ${\bf 3}$ but with the constraint $\textstyle{\sum_i}|e^{(i)}_{\bf M}\rangle=0$ (this constraint gets rid of ${\bf 1_S}\in {\bf 3}$ leaving us with ${\bf 2_M}$), these symmetric generators are,
\be
s|e^{(1)}_{\bf M}\rangle+t|e^{(2)}_{\bf M}\rangle+u|e^{(3)}_{\bf M}\rangle\quad{\rm and}\quad tu|e^{(1)}_{\bf M}\rangle+us|e^{(2)}_{\bf M}\rangle+st|e^{(3)}_{\bf M}\rangle.
\ee
Note that this means that, for purposes of constructing physical S-matrix, a ${\bf 2_M}$ generator with derivative order $\partial$ is equivalent to two ${\bf 1_S}$ generators at derivative orders $\partial+2$ and $\partial+4$. This is what the formula in \eqref{Z-S3} reflects.
The S-matrix from generator transforming in ${\bf 1_A}$ is constructed in the same way i.e. first constructing a symmetric generator and multiplying it by symmetric polynomials of $(s,t)$. If we take $|e_{\bf A}\rangle$ to be the generator in ${\bf 1_A}$, the associated symmetric generator is,
\be
s^2u-u^2s+t^2s-s^2t+u^2t-t^2u \,|e_{\bf A}\rangle.
\ee
Note that this means that a ${\bf 1_A}$ generator with derivative order $\partial$ is equivalent to a ${\bf 1_S}$ generator at derivative order $\partial+6$. This is what the formula in \eqref{Z-S3} reflects.

Now, the computation of the partition function over physical S-matrices following from a set of quasi-invariant generators and relations is clear.
\be\label{pf-from-gen}
Z_{\rm S-matrix}=\frac{1}{(1-x^4)(1-x^6)}\Big(\sum_{e\in \bf S} x^{\partial_e}+\sum_{e\in \bf M} (x^2+x^4)\, x^{\partial_e}+\sum_{e\in \bf A}x^6\, x^{\partial_e}\Big).
\ee
Relations contribute to the above sums with a negative sign. However, note that the partition function when expanded in powers of $x$ must have positive integer coefficients.

\subsubsection{Constructing gluon module from photons and adjoint scalars}

The modules $\CM^{\rm gluon}, \CM^{\rm photon}$ and $\CM^{\rm scalar}$ of gluon, photon and adjoint scalar quasi-invariant S-matrices respectively are the main players of our game. Among these, $\CM^{\rm photon}$ has been characterized in detail in \cite{Chowdhury:2019kaq}. In this paper, our goal is to characterize $\CM^{\rm gluon}$. In order to do so we will first characterize $\CM^{\rm scalar}$ and use the tensor product structure \eqref{tensor} between photon and scalars to characterize $\CM^{\rm gluon}$. In this section, we will describe how this tensor product works. Before that let us comment a bit on $\CM^{\rm scalar}$. An important thing to note about this module is that all its generators have $0$ derivative order and that they generate the module freely. This is because, the momenta $p_i$ appear in the scalar S-matrix only through Mandelstam variables. This is unlike photon S-matrix where they can dot with polarization vectors. Hence, the generators of the module must not have any momenta in them.  This argument also holds for relations, if any. Because, the generators and relations both appear at $0$ derivative, the relations can simply be removed from the set of generators. The remaining ones generate $\CM^{\rm scalar}$ freely. These generators are nothing but color structures. Sometimes it is convenient to think of them as forming a vector space $\CV^{\rm scalar}$ over ${\mathbb C}$. 

From equation \eqref{tensor},  it is tempting to surmise,
\be
\CM^{\rm gluon}=\CM^{\rm photon} \otimes \CV^{\rm scalar}.
\ee
The tensor product above certainly yields quasi-invariant gluon S-matrices but this is not the whole module $\CM^{\rm gluon}$. The quasi-invariant gluon S-matrix could also come from the tensor product of photon and scalar pieces that are not separately quasi-invariant. Let the module of such gluon S-matrices be $\CM^{\rm non-inv}$. 
The correct description of the gluon module is then,
\bea
\CM^{\rm gluon}&=& \CM^{\rm inv} \oplus \CM^{\rm non-inv},\nonumber\\
\CM^{\rm inv} &\equiv & \CM^{\rm photon} \otimes \CV^{\rm scalar}.
\eea

\subsection{Outline}
We now describe the outline of the rest of the paper. In section \ref{s3-non-inv} we describe how to assign $S_3$ representation to the non-quasi-invariant states and how to take the tensor product of two such representations and project onto quasi-invariant states. This is used in the construction of $\CM^{\rm non-inv}$ in section \ref{non-inv-tensor}. 
In section \ref{counting} we construct formulas for counting the dimension quasi-invariant color structures i.e. of $\CV^{\rm scalar}$ and also for counting of the non-quasi-invariant color structure. We keep track of the $S_3$ representation in both cases and the results are listed in table \ref{quasi-inv-counting}.
In section \ref{plethystic}, we  compute the partition function over $\CM^{\rm scalar}$ for all dimensions 
and over $\CM^{\rm gluon}$ for all dimensions and for gauge group $G=SO(N)$ and $SU(N)$. The main results of this section are tabulated in tables \ref{scalar-plethystic} for $\CM^{\rm scalar}$ and tables \ref{gluon-plethystic}, \ref{gluonplethd7}, \ref{gluonplethd6}, \ref{gluonplethd5} and  \ref{gluonplethd4} for $\CM^{\rm gluon}$. The partition function over $\CM^{\rm scalar}$, using the relation in section \ref{partition}, serves as a check for the basis of $\CV^{\rm scalar}$ enumerated in section \ref{counting} . The partition function over $\CM^{\rm gluon}$ is our main check over the explicit construction of gluon S-matrices in section \ref{explicit}.
In the last section \ref{explicit}, we explicitly construct $\CM^{\rm {gluon}}$ by first constructing the $\CM^{\rm inv}$ and then constructing $\CM^{\rm non-inv}$. In both cases, we first consider the photon part and the scalar part separately and then take their tensor product. The tensor product also needs to be projected onto non-quasi-invariant states in the case of  $\CM^{\rm non-inv}$ which we do. We obtain an explicit description of all the local Lagrangians which generate the local gluon S-matrices in asymptotically high dimensions (i.e $D \geq 9$) and for gauge group $G=SO(N)$ and $SU(N)$ for all $N$.  Although we do not list them explicitly, the gluon S-matrices
for the lower dimensions can be similarly worked out. In section \ref{summary}, we summarize the results and end with outlook. The paper is supplemented with four appendices.

\section{Non-quasi-invariants}\label{s3-non-inv}

Because, $\Z_2\times \Z_2$ appears as a kernel of a representation, it is a normal subgroup of $S_4$. The quotient $S_4/(\Z_2\times \Z_2)$ is the group $S_3$, alternatively, $S_4$ is  the semi-direct product $S_3 \ltimes (\Z_2\times \Z_2)$. This means that $S_3$ acts non-trivially on $\Z_2\times \Z_2$. This can be seen in the list of elements of $\Z_2\times \Z_2$ given in  \eqref{v4list}. The elements get permuted under $S_3$ i.e. under the permutations of $2,3,4$.
This action lifts to the $S_3$ action on $\Z_2\times \Z_2$ representations. 

The $\Z_2\times \Z_2$ is an abelian group. It's representations are specified by giving the charges under its generators $(P_{12}P_{34},P_{13}P_{24})$. For convenience, we will also denote the charge under their product $P_{14}P_{23}$ and label the representation by $(P_{12}P_{34},P_{13}P_{24},P_{14}P_{23})$. There are four irreducible representations, all of them one-dimensional. They are given be the charges $(+,+,+),\,(+,-,-),\,(-,+,-)$ and $(-,-,+)$. Among these, $(+,+,+)$ is quasi-invariant while the other three are non-quasi-invariant. It is not difficult to see that the action of $S_3$ relates the non-quasi-invariant representations. This is as follows.

 Let $|\psi\rangle$ be a non-invariant representation of $\Z_2\times \Z_2$. Let us look at the $P_{12}P_{34}$ charge of the state $\sigma|\psi\rangle$ where $\sigma\in S_3$.
\be
P_{12}P_{34} (\sigma|\psi\rangle)=\sigma(\sigma^{-1} P_{12}P_{34} \sigma)|\psi\rangle=g P_{\sigma(1)\sigma(2)} P_{\sigma(3)\sigma(4)}|\psi\rangle= P_{\sigma(1)\sigma(2)} P_{\sigma(3)\sigma(4)}(\sigma|\psi\rangle).
\ee
Here $\sigma|i\rangle$ denotes the image of $i$ under the action of $\sigma$. If $\sigma=P_{12}$ or $P_{34}$ then $P_{\sigma(1)\sigma(2)} P_{\sigma(3)\sigma(4)}=P_{12}P_{34}$ and for any other choice of $\sigma$, $\sigma|\psi\rangle$ gets its $\Z_2\times \Z_2$ charges permuted. Now let $|\psi_1\rangle$ be the state with charge $(+--)$. The state $|\psi_2\rangle\equiv P_{12}|\psi_1\rangle$ also has the same charges. The $6$ dimensional orbit of state $|\psi_1\rangle$ under the action of $S_3$ elements splits into two $3$ dimensional orbits of $|\psi_1\rangle\pm|\psi_2\rangle$ respectively. The orbit of $|\psi_1\rangle +|\psi_2\rangle$  forms the representation ${\bf 3}={\bf 1_S}+{\bf 2_M}$ of $S_3$ while the orbit of $|\psi_1\rangle -|\psi_2\rangle$ forms the representation ${\bf 3_A}={\bf 1_A}+{\bf 2_M}$ of $S_3$.

\subsection{Tensor product and projection}

Let $|e^{(1),(2),(3)}\rangle$  be the three terms of the photon S-matrix with $\Z_2\times \Z_2$ charges $(+--,-+-,--+)$ respectively. As the $S_3$ representation, it could transform either as ${\bf 3}$ or as ${\bf 3_A}$. Let $|f^{(1),(2),(3)}\rangle$ be the three color structures of adjoint scalars with $\Z_2\times \Z_2$ charges $(+--,-+-,--+)$ respectively. Again, it could transform either as ${\bf 3}$ or as ${\bf 3_A}$ under $S_3$. From these states,  quasi-invariant module generators can be constructed as
\be
|e^{(1)}f^{(1)}\rangle, |e^{(2)}f^{(2)}\rangle, |e^{(3)}f^{(3)}\rangle.
\ee
Note that these are the only quasi-invariant states in the tensor product of $|e^{(i)}\rangle$ and $|f^{(i)}\rangle$. The $S_3$ representation that it forms depends on the $S_3$ representation of $|e^{(i)}\rangle$ and $|f^{(i)}\rangle$. If $|e^{(i)}\rangle$ and $|f^{(i)}\rangle$ are both ${\bf 3}$ or both ${\bf 3_A}$ then the quasi-invariant module generator $|e^{(i)}f^{(i)}\rangle$ forms the representation ${\bf 3}$. If one of $|e^{(i)}\rangle$ and $|f^{(i)}\rangle$ is ${\bf 3}$ and the other is ${\bf 3_A}$ then $|e^{(i)}f^{(i)}\rangle$ forms the representation ${\bf 3_A}$. This follows simply by observing that the $P_{12}$ charge of $|e^{(1)}f^{(1)}\rangle$ is the product of $P_{12}$ charge of $|e^{(1)}\rangle$ and $|f^{(1)}\rangle$.
We expect the module $\CM^{\rm non-inv}$ to be freely generated by the states of the type $|e^{(i)}f^{(i)}\rangle$. It is a free module because different color structure do not get related after  multiplication by the ring elements i.e. polynomials of the Mandelstam variables $(s, t)$ if they are already not related.

\section{Counting colour modules using projectors}\label{counting}
In this subsection, we will develop a group theory formula that gives the $G$-representation in the quasi-invariant sector as well as non-quasi-invariant sector of the tensor product of four identical $G$-representations along with their $S_3$ representations. This is done by constructing a projector over states with appropriate $\Z_2\times \Z_2$ charge. Further projecting this $G$-representation onto $G$-singlets gives the number of quasi-invariant and non-quasi-invariant color structures. 
This counting serves as a check over the explicit construction of the color structures. 

Before moving to the counting problem of interest, let us quickly review the construction of symmetric (anti-symmetric) representation in the tensor product of two identical $G$-representations $\rho$.
Let $|\alpha\rangle$ denote the basis vectors of $\rho$. The character is
\be
\chi_\rho(a_i)=\sum_\alpha \langle \alpha | \prod_i a_i^{H_i} |\alpha\rangle.
\ee
In the states in the tensor product of two such representations is $|\alpha_1,\alpha_2\rangle$. The projector onto the symmetric(anti-symmetric) states is $(1\pm P_{12})/2$. The character of the representation in the symmetric product is
\bea
\chi_{S^2\rho\,\,(\wedge^2\rho)}(a_i)&=&\sum_{\alpha_1 \alpha_2} \langle \alpha_1,\alpha_2 | a_i^{H_i} \,\frac{1\pm P_{12}}{2} |\alpha_1,\alpha_2\rangle,\nonumber\\
&=& \frac12 \Big(\sum_{\alpha_1} \langle \alpha_1 | \prod_i a_i^{H_i} |\alpha_1\rangle\Big)\Big(\sum_{\alpha_2} \langle \alpha_2 | \prod_i a_i^{H_i} |\alpha_2\rangle\Big)\nonumber\\
&\pm &\frac12 \sum_{\alpha_1 \alpha_2} \langle \alpha_1,\alpha_2 | a_i^{H_i}  |\alpha_2,\alpha_1\rangle.
\eea
Note that  ${\textstyle \sum_{\alpha_2}}\langle \alpha_1,\alpha_2 | a_i^{H_i}  |\alpha_2,\alpha_1\rangle={\textstyle \sum_{\alpha_2}}\langle\alpha_1 | \prod_i a_i^{H_i} |\alpha_2\rangle\langle\alpha_2 | \prod_i a_i^{H_i} |\alpha_1\rangle=\langle \alpha_1 | \prod_i (a_i^2)^{H_i} |\alpha_1\rangle$. Using this,
\be
\chi_{S^2\rho \,\,(\wedge^2\rho)}(a_i)=\frac12(\chi_\rho(a_i)^2\pm \chi_\rho(a_i^2)).
\ee

\subsection{Counting of quasi-invariant colour modules}\label{count-inv}

The quasi-invariant i.e. $\Z_2\times \Z_2$ invariant projector is constructed in the same way. On the tensor product of four copies of representation $\rho$, the $\Z_2\times \Z_2$ invariant projector is,
\be
\Big(\frac{1+P_{12}P_{34}}{2}\Big)\Big( \frac{1+P_{13}P_{24}}{2}\Big) \Big(\frac{1+P_{14}P_{23}}{2} \Big)=\frac14 (1+P_{12}P_{34}+P_{13}P_{24}+P_{14}P_{23}).
\ee
In this line, we have used the fact that $P_{ij}^2=1$ and $(P_{12}P_{34})(P_{13}P_{24})=P_{14}P_{23}$. The quasi-invariant part of the tensor product is \cite{Kravchuk:2016qvl},
\be\label{z2z2invk}
{\rho^{\otimes 4}|_{\Z_2\times\Z_2}}= \rho^{\otimes 4} - 3(S^2\rho \otimes \wedge^2 \rho).
\ee
 In order to get the number of quasi-invariant color structures we project this $G$-representation onto singlets. However, this provides us only with the total number of quasi-invariant color structures and doesn't give information about their $S_3$ representations. If the quasi-invariant states consist of $n_{\bf S}$ number of ${\bf 1_S}$, $n_{\bf M}$ number of ${\bf 2_M}$ and $n_{\bf A}$ number of ${\bf 1_A}$. Then 
\be\label{proj1}
n_{\bf S}+2n_{\bf M}+n_{\bf A}=\rho^{\otimes 4} - 3(S^2\rho \otimes \wedge^2 \rho)|_{\bf G}.
\ee
In order to count the number of $S_3$ representations separately we need two more equations. They are obtained as follows.

The quasi-invariant states that transform in ${\bf 1_S}$ are the states that are invariant under the entire $S_4$. This means
\be\label{proj2}
n_{\bf S}=S^4\rho|_{G}.
\ee
We can also count the number of ${\bf 1_A}$ by constructing a projector onto states that are invariant under $\Z_2\times \Z_2$ but are antisymmetric under $S_3$. This projector is
\be\label{1az2z2}
\frac{\left(1-P_{12}-P_{23}-P_{13}+P_{23}P_{12}+P_{12}P_{23}\right)}{6}\frac{(1+P_{12}P_{34}+P_{13}P_{24}+P_{14}P_{23})}{4}.
\ee
From this projector we obtain the $G$-representation that is in ${\bf 1_A}$.
\be\label{proj3}
n_{\bf A}=(\wedge^2\rho)^2 - \wedge^2\rho \otimes S^2\rho - (S^2\rho)^2 - S^4\rho + 2 S^3\rho \otimes \rho|_{G}
\ee
See appendix \ref{projectors} for the details of this computation. Using similar techniques one can verify eq \eqref{z2z2invk}. From equations \eqref{proj1},\eqref{proj2} and \eqref{proj3}, the values of $n_{\bf S}, n_{\bf M}$ and $n_{\bf A}$ can be  obtained. We have computed them for $G=SO(N)$ and $G=SU(N)$ and tabulated the result in table \ref{quasi-inv-counting}. Counting for other Lie groups is straightforward.
\begin{table}
		\begin{center}
			\begin{tabular}{|l|l|l|l|}
				\hline
				$SO(N)$ & $n_{\bf S}$ & $n_{\bf M}$ & $n_{\bf A}$ \\
				\hline
				$N\geq  9$ & 2& 2 &0\\
				\hline
				$N=8$ & 3 & 2& 0\\
				\hline
				$N=4$ & 3 & 3 & 0 \\
				\hline
			\end{tabular}\\
			\vspace{0.3cm}
			\begin{tabular}{|l|l|l|l|}
				\hline
				$SU(N)$ & $n_{\bf S}$ & $n_{\bf M}$ & $n_{\bf A}$ \\
				\hline
				$N\geq  4$ & 2& 2 &0\\
				\hline
				$N=3$ & 1 & 2& 0\\
				\hline
				$N=2$ & 1 & 1 & 0\\
				\hline
			\end{tabular}
		\end{center}
		\caption{The counting of $S_3$ representations of quasi-invariant color structures. The results for $SO(N)$ with $N=7,6,5$ are the same as those for $N\geq 9$.}
		\label{quasi-inv-counting}
\end{table}

Note that the counting of the four-point color structures is independent of $N$ for $SO(N\geq 9)$ and for $SU(N\geq 4)$. This can be understood using a variation of the argument presented in \cite{Chowdhury:2019kaq} to show that the counting of four-point photon S-matrices for $D\geq 8$ is independent of $D$.
Let us review that argument here. 
The scattering of four particles takes place in a three dimensional space. The transverse polarization  of the first photon points in the fourth direction. Then the transverse polarizations of the 2nd, 3rd and 4th photon can generically be taken to point in 5th, 6th and 7th direction and any additional dimensions simply play the role of the spectator. Hence for $D\geq 8$, the counting of photon S-matrices is independent of $D$.\footnote{In $D=7$ a single parity odd S-matrix exists because 7 vectors involved in scattering (4 polarizations and 3 momenta) can be contracted by a 7 index Levi-Civita tensor but for the same reason, no parity odd S-matrix exists for $D\geq 8$.} This argument also tells us that the counting of four-point gluon S-matrices is also independent of $D$ for $D\geq 8$.

For scalars transforming in the adjoint (anti-symmetric) representation of $SO(N)$, each scalar labels a plane in the internal $N$-dimensional space. Four such planes generically span 8-dimensional subspace of the $N$-dimensional space. Any additional transverse dimensions of the internal space are spectators. Hence the counting is independent of $N$ for $SO(N\geq 9)$.\footnote{In the case of $SO(8)$, the parity odd (in the internal space) structure is present but parity odd structures are absent for $N\geq 9$.} For $SU(N)$ adjoint scalars, the independence with respect to $N$ for $N\geq 4$ can be understood  using the fact that four $SU(N)$ matrices are always independent for $N\geq 4$, but for $N\leq 3$ the four $SU(N)$ matrices obey relations.

\subsection{Counting of non-quasi-invariant color modules}\label{count-non-inv}

We also count the number of non-quasi-invariant color structures that transforms in ${\bf 3}$ (${\bf 3_A}$) using an appropriate projector.  From the discussion in section \ref{s3-non-inv},  it is clear that the number of such representations is the same as the number of states that has the charges $(+--)$ under $\Z_2\times \Z_2$ and are even (odd) under the action of $P_{12}$. The projector on such states is
\be
\Big(\frac{1\pm P_{12}}{2}\Big)\Big(\frac{1+P_{12}P_{34}}{2}\Big)\Big( \frac{1-P_{13}P_{24}}{2}\Big) \Big(\frac{1-P_{14}P_{23}}{2} \Big)=\Big(\frac{1\pm P_{12}}{2}\Big)\frac14 (1+P_{12}P_{34}-P_{13}P_{24}-P_{14}P_{23}).
\ee
Again we have used, $P_{ij}^2=1$ and $(P_{12}P_{34})(P_{13}P_{24})=P_{14}P_{23}$. The representation of such states in the tensor product of four copies of representation $\rho$ is,
\bea\label{proj4}
\rho_{{\bf 3}}&=& -S^4\rho + S^3\rho \otimes \rho,\\
\rho_{{\bf 3_A}}&=& S^4\rho - S^3\rho \otimes \rho+ \wedge^2\rho \otimes S^2\rho.\nonumber
\eea
The computation of this group theory formula is detailed in appendix \ref{projectors}. The number of color structures transforming in ${\bf 3}$ (${\bf 3_A}$) representation is obtained by projecting the above representation onto $G$-singlets. The result for $G=SO(N)$ and $G=SU(N)$ is tabulated in \ref{non-quasi-inv-counting}. Counting for other Lie groups is straightforward.
\begin{table}
		\begin{center}
			\begin{tabular}{|l|l|l|}
				\hline
				$SO(N)$ & $n_{\bf 3}$ & $n_{\bf 3_A}$  \\
				\hline
				$N\geq  7$ & 0& 0 \\
				\hline
				$N=6$ & 0 & 1  \\
				\hline
				$N=4$ & 1 & 0\\
				\hline
			\end{tabular}\\
			\vspace{0.3cm}
			\begin{tabular}{|l|l|l|}
				\hline
				$SU(N)$ & $n_{\bf 3}$ & $n_{\bf 3_A}$ \\
				\hline
				$N\geq  3$ & 0& 1 \\
				\hline
				$N=2$ & 0 & 0 \\
				\hline
			\end{tabular}
		\end{center}
		\caption{The counting of $S_3$ representations of non-quasi-invariant color structures. The results for $SO(N)$ with $N=5,3$ are the same as those for $N\geq 8$.}
		\label{non-quasi-inv-counting}
\end{table}

In the next section, we will explicitly construct the quasi-invariant and quasi-non-invariant color structures  and check their number against the counting in tables \ref{quasi-inv-counting} and \ref{non-quasi-inv-counting}. We will also explicitly construct non-quasi-invariant photon S-matrices.

\section{Counting gluon Lagrangians with plethystics}\label{plethystic}
As emphasized in \cite{Chowdhury:2019kaq}, the space of S-matrices of a theory is isomorphic to the Lagrangians of the theory modulo equations of motion and total derivative. In this section, we will use plethystic techniques to derive an integral formula to enumerate Yang-Mills Lagrangians module equations of motion and total derivatives thereby constructing the partition function over the space of S-matrices. Similar techniques were used in \cite{Henning:2015daa} to compute the partition function on ``operator bases" in effective field theories and were generalized to compute partition function on S-matrices in \cite{Henning:2017fpj}.
They were also used in \cite{Sundborg:1999ue, Aharony:2003sx} to compute the partition function of gauge theories at weak coupling.

As a warm-up we will do this for $G$-adjoint scalars. From the resulting partition function over the S-matrices, the number quasi-invariant module generators are read off. This matches the quasi-invariant color structures counted using a group theory formula in section \ref{count-inv} and explicitly written down in section \ref{quasi-inv-color}. 

\subsection{Scalars}\label{pleth-scalar}
We first evaluate the single letter partition function over the scalar in representation $R$ of a global symmetry. Later, we will specialize the representation $R$ to adjoint. The space of single letter operators is spanned by
\be
\partial_{\mu_1}\ldots \partial_{\mu_l} \Phi^a,\qquad l\geq 0.
\ee
The equation of motion is $\partial_\mu\partial^\mu \Phi^a$. Here $a$ is the internal index. The single letter index is
\begin{eqnarray}\label{scalar-single}
i_\ts(x,y,z)&=&{\rm Tr}\,\,x^{\partial} y_i^{L_i}z_\alpha^{H_\alpha}= \chi_R(z)(1-x^2)\denom(x,y) \equiv \chi_R(z)i_\ts(x,y) .\nonumber\\
\denom(x,y) &=&\Big(\prod_{i=1}^{D/2}(1-x y_i)(1-x y_i^{-1})\Big)^{-1}\qquad \qquad \qquad \qquad{\rm for \,\, D\,\, even}\nonumber \\
&=&\Big((1-x)\prod_{i=1}^{\lfloor D/2\rfloor}(1-x y_i)(1-x y_i^{-1})\Big)^{-1}\qquad \qquad \,\,{\rm for \,\, D\,\, odd}.
\end{eqnarray}
Here $\partial$ is the number of derivatives, $L_i$ and $H_\alpha$ stand for the Cartan elements of $SO(D)$ and $G$ respectively. The denominator factor $\denom(x,y)$ encodes the tower of derivatives on $\Phi(x)$ keeping track of the degree and the charges under the Cartan subgroup of $SO(D)$. The factor $\chi_R(z)$ is the character of the representation $R$. We will eventually project onto $G$-singlets.

The partition function over four identical scalars, relevant for counting quartic Lagrangians, is given by,
\bea\label{4-particle}
i_\ts^{(4)}(x,y,z)&=&\frac{1}{24}\Big(i^4_\ts(x,y,z) +6 i^2_\ts(x,y,z) i_\ts(x^2,y^2,z^2)+3i^2_\ts(x^2,y^2,z^2)\nonumber\\
&+&8i_\ts(x,y,z)i_\ts(x^3,y^3,z^3)+6i_\ts(x^4,y^4,z^4)\Big).\nonumber\\
&=&\frac{1}{24}\Big(i^4_\ts(x,y)~\chi_R^G(z)^4 +6 i^2_\ts(x,y) i_\ts(x^2,y^2)~\chi_R^G(z)^2\chi_R^G(z^2)+3i^2_\ts(x^2,y^2)~\chi_R^G(z^2)^2\nonumber\\
&+&8i_\ts(x,y)i_\ts(x^3,y^3)~\chi_R^G(z)\chi_R^G(z^3)+6i_\ts(x^4,y^4)~\chi_R^G(z^4)\Big).
\eea
The Lagrangians that are total derivatives can be removed by dividing by ${\tt D}(x,y)$.
Now to construct Lorentz invariant and $G$-invariant Lagrangians we integrate over $SO(D)$ and $G$ with the Haar measure.  
\be\label{singlet-proj}
I^D_\ts(x):=\oint  d\mu_{G}~ \oint  d\mu_{SO(D)}~  i_\ts^{(4)}(x,y,z)/\denom(x,y).
\ee  
Here $d\mu_{G}$ is the Haar measure associated with the group $G$.

Using this formula the partition function over four-point S-matrix of scalars transforming in representation $R$ of $G$ in dimension $D$ can be constructed. We will do the integral and tabulate the result for $D\geq 3$ and for $G=SO(N)$ and $SU(N)$. In dimensions $D\geq 3$, the partition function is independent of $D$, so $D$ can be taken to be large and integral can be performed by saddle point method. After doing the $SO(D)$ integral in the $D\to \infty$ limit,
\be\label{scalar-proj-colour}
\begin{split}
I^D_\ts(x):= \oint d\mu_{G}~&\left(\frac{\chi^{G}_R(z^2) \chi^{G}_R(z)^2}{4 \left(1-x^4\right)}+\frac{\chi^{G}_R(z^4)}{4 \left(1-x^4\right)}+\frac{\chi^{G}_R(z)^4}{24 \left(x^2-1\right)^2}+\frac{\chi^{G}_R(z^2)^2}{8 \left(x^2-1\right)^2}\right.\nonumber\\
&\left. +\frac{\chi^{G}_R(z^3)\chi^{G}_R(z)}{3 \left(x^4+x^2+1\right)}\right).
\end{split}
\ee 
 For adjoint representation of $G=SO(N)$, the colour integral \eqref{scalar-proj-colour} stabilizes for $N\geq 9$, while for adjoint of $G=SU(N)$, it stabilizes for $N\geq 4$. For large $N$, we perform the integral using Large $N$ techniques outlined in appendix \ref{pi} (see subsection \ref{pison} for $SO(N)$ and subsubsection \ref{pisunlargen} for $SU(N)$ large $N$ integrals). The integrals for $N<9$ for $SO(N)$ and $N<4$ for $SU(N)$ have been done numerically using the methods outlined in appendix \ref{pison} and \ref{pisunsmalln} respectively. The result of this integral for the adjoint representation of $SO(N)$ and $SU(N)$ is tabulated in table \ref{scalar-plethystic}.
\begin{table}
		\begin{center}
			\begin{tabular}{|l|l|}
				\hline
				$SO(N)$ & scalar partition function \\
				\hline
				$N\geq  9$ & $({2+2x^2+2x^4})\denom$\\
				\hline
				$N=8$ & $({3+2x^2+2x^4})\denom$\\
				\hline
				$N=4$ & $({3+3x^2+3x^4})\denom$\\
				\hline
				$N=3$ & $({1+x^2+x^4})\denom$ \\
				\hline
			\end{tabular}\\ \vspace{0.3cm}
			\begin{tabular}{|l|l|}
				\hline
				$SU(N)$ & scalar partition function \\
				\hline
				$N\geq  4$ & $({2+2x^2+2x^4})\denom$\\
				\hline
				$N=3$ & $({1+2x^2+2x^4})\denom$\\
				\hline
				$N=2$ & $({1+x^2+x^4})\denom$\\
				\hline
			\end{tabular}
		\end{center}
		\caption{Partition function over the space of Lagrangians involving four $\Phi^a$'s that are in the adjoint of $SO(N)$ and $SU(N)$. This includes both parity even and parity odd Lagrangians. We have defined $\denom\equiv 1/((1-x^4)(1-x^6))$.}
		\label{scalar-plethystic}
\end{table}
This counting agrees with the counting of color structures in table \ref{quasi-inv-counting}.

Note that a complete classification of scalar field primaries transforming in any representation and with arbitrary number of $\Phi$'s has been carried out using algebraic methods in \cite{deMelloKoch:2017dgi, deMelloKoch:2018klm}.

\subsection{Gluons}\label{pleth-gluon}
The partition function over four-gluon S-matrices is computed in the same way. We first construct the single letter partition function. This is same as the single letter partition function for photons multiplied by the adjoint character. Using the single letter photon partition function computed in \cite{},
\begin{eqnarray}\label{gluon-single}
i_\tv(x,y,z)&=&{\rm Tr}\,\,x^{\partial} y_i^{L_i}z_\alpha^{H_\alpha}= \chi_{\rm adj}(z)(((x-x^3)\chi_{\syng{1}}-(1-x^4))\denom(x,y)+1)/x \equiv \chi_{\rm adj}(z) i_\tv(x,y)\nonumber\\
\chi_{\syng{1}} &=& \sum_{i=1}^{D/2} \left( y_i + \frac{1}{y_i}\right)\qquad \qquad \qquad \qquad{\rm for \,\, D\,\, even}\nonumber \\
&=& \sum_{i=1}^{\lfloor D/2 \rfloor} \left( y_i + \frac{1}{y_i}\right)+1\qquad \qquad \qquad {\rm for \,\, D\,\, odd}.
\end{eqnarray}
The four particle partition function is, 
\bea\label{4-particlegluon}
	i_\tv^{(4)}(x,y,z)&=&\frac{1}{24}\Big(i^4_\tv(x,y,z) +6 i^2_\tv(x,y,z) i_\tv(x^2,y^2,z^2)+3i^2_\tv(x^2,y^2,z^2)+8i_\tv(x,y,z)i_\tv(x^3,y^3,z^3)\nonumber \\
	&+&6i_\tv(x^4,y^4,z^4)\Big)\nonumber\\
	&=&\frac{1}{24}\Big(i^4_\tv(x,y)~\chi_R^G(z) +6 i^2_\tv(x,y) i_\tv(x^2,y^2)~\chi_R^G(z)^2\chi_R^G(z^2)+3i^2_\tv(x^2,y^2)~\chi_R^G(z^2)^2\nonumber\\
	&+&8i_\tv(x,y)i_\tv(x^3,y^3)~\chi_R^G(z)\chi_R^G(z^3)+6i_\tv(x^4,y^4)~\chi_R^G(z^4)\Big).
\eea  
Getting rid of the total derivatives by dividing by $\denom(x,y)$ and projecting onto singlets of $SO(D)$ and $G$ by integrating over their Haar measures,
\be\label{singlet-projgluon}
I^D_\tv(x):=\oint  d\mu_{G}~ \oint  d\mu_{SO(D)}~  i_\tv^{(4)}(x,y,z)/\denom(x,y).
\ee  
We will do the integral and tabulate the result for $D\geq 8$ and for $G=SO(N)$ and $SU(N)$. In dimensions $D\geq 8$, the partition function is independent of $D$, so $D$ can be taken to be large and integral can be performed by saddle point method. After doing the $SO(D)$ integral in the $D\to \infty$ limit,
\bea
	I_\tv^{D}(x) &=&\oint  d\mu_{G}~ \left(\frac{x^8 \chi^{R}_a(z^4)}{2-2 x^4} +\frac{\left(x^8 \left(2 x^2+3\right)\right) \chi^{R}_a(z)^4}{12 \left(x^2-1\right)^2}+\frac{\left(3 x^8\right) \chi^{R}_a(z^2)^2}{4 \left(x^2-1\right)^2} \right.\nonumber\\
	&&\left.~~~~~~~~~~ +\frac{x^8 \chi^{R}_a(z)^2 \chi^{R}_a(z^2)}{2-2 x^2}+\frac{x^{10} \chi^{R}_a(z) \chi^{R}_a(z^3)}{3 \left(x^4+x^2+1\right)}\right).
\eea 
The result of this integral for $G=SO(N)$ and $SU(N)$ is tabulated in \ref{gluon-plethystic}. We resort to numerical integration for $D \leq8$ and for $SO(N\leq 8)$ and $SU(N\leq 3)$ partition function evaluation. 
The results for $D=7$, $D=6$, $D=5$ and $D=4$ are presented in tables \ref{gluonplethd7}, \ref{gluonplethd6}, \ref{gluonplethd5} and \ref{gluonplethd4} respectively. In all the cases, the contribution of ${\CM}^{\rm inv}$ is identified using the explicit construction of $\CM^{\rm inv}$ in section \ref{M-inv}. The rest of the contribution is attributed to $\CM^{\rm non-inv}$ and is underlined.
\begin{table}
		\begin{center}
			\begin{tabular}{|l|l|}
				\hline
				$SO(N)$ & Gluon partition function  \\
				\hline
				$N\geq  9$ & ${2x^4(4+7x^2+7x^4+3x^6)}\denom$\\
				\hline
				$N=8$ & ${(2x^4(5+8x^2+8x^4+3x^6)+x^6)}\denom$ \\
				\hline
				$N=6$ & $({2x^4(4+7x^2+7x^4+3x^6)}+{\underline {x^6(x^2+x^4+x^6)}})\denom$ \,\\
				\hline
				$N=4$ & $(3x^4(4+7x^2+7x^4+3x^6)+{\underline{x^6(1+x^2+x^4)}})\denom$ \\
				\hline
			\end{tabular}\\
			\vspace{0.3cm}
			\begin{tabular}{|l|l|}
				\hline
				$SU(N)$ & Gluon partition function  \\
				\hline
				$N\geq  4$ & $(2x^4(4+7x^2+7x^4+3x^6)+\underline{x^6(x^2+x^4+x^6)})\denom$\\
				\hline
				$N=3$ & $(x^4(6+11x^2+12x^4+6x^6)+\underline{x^6(x^2+x^4+x^6)})\denom$\\
				\hline
				$N=2$ & $x^4(4+7x^2+7x^4+3x^6)\denom$ \\
				\hline
			\end{tabular}
		\end{center}
		\caption{Partition function over the space of Lagrangians in $D\geq 8$ involving four $F^a_{\a\b}$'s. Recall $\denom\equiv 1/((1-x^4)(1-x^6))$. Contribution of $\CM^{\rm non-inv}$ is underlined. The rest is the contribution of $\CM^{\rm inv}$. Partition function for $SO(5)$ and $SO(7)$ is the same as that for $SO(N)$ for $N\geq 9$.}
		\label{gluon-plethystic}
	\end{table}

\begin{table}
	\begin{center}
		\begin{tabular}{|l|l|}
			\hline
			$SO(N)$ & Gluon partition function   \\
			\hline
			$N\geq 9$ & $  2x^3(1 + 4 x + x^2 + 7 x^3 + x^4 + 7 x^5 + 3 x^7)\denom -2x^3$\\
			\hline
			$N = 8$ & $ x^3 (3 + 10 x + 2 x^2 + 17 x^3 + 2 x^4 + 16 x^5 + 6 x^7)\denom -3x^3$\\
			\hline
			$N=6$ & $ 2x^3(1 + 4 x + x^2 + 7 x^3 + x^4 + 7 x^5 + 3 x^7)\denom -2x^3+ {\underline{x^6(x^2+x^4+x^6)\denom}}\quad$ \\
			\hline
			$N= 4$ &$ 3x^3(1 + 4 x + x^2 + 7 x^3 + x^4 + 7 x^5 + 3 x^7)\denom -3x^3+ {\underline{x^6(1+x^2+x^4)\denom}}$\\
			\hline
		\end{tabular}\\
		\vspace{0.3cm}
		\begin{tabular}{|l|l|}
			\hline
			$SU(N)$ & Gluon partition function  \\
			\hline
			$N\geq 4$ & $ 2x^3(1 + 4 x + x^2 + 7 x^3 + x^4 + 7 x^5 + 3 x^7)\denom -2x^3+ {\underline{x^6(x^2+x^4+x^6)\denom}}$ \\
			\hline
			$N=3$ & $x^3 (1 + 6 x + 2 x^2 + 11 x^3 + 2 x^4 + 12 x^5 + 6 x^7)\denom-x^3+ \underline{x^6(x^2+x^4+x^6)\denom}$\\
			\hline
			$N=2$ & $x^3(1 + 4 x + x^2 + 7 x^3 + x^4 + 7 x^5 + 3 x^7)\denom -x^3$\\
			\hline
		\end{tabular}
	\end{center}
	\caption{Partition function over the space of Lagrangians in $D=7$ involving four $F^a_{\a\b}$'s. Recall $\denom\equiv 1/((1-x^4)(1-x^6))$. Contribution of $\CM^{\rm non-inv}$ is underlined. The rest is the contribution of $\CM^{\rm inv}$. Partition function for $SO(5)$ and $SO(7)$ is the same as that for $SO(N)$ for $N\geq 9$.}
	\label{gluonplethd7}
\end{table}

\begin{table}
	\begin{center}
		\begin{tabular}{|l|l|}
			\hline
			$SO(N)$ & Gluon partition function  \\
			\hline
			$N\geq 9$ & $2 x^4 (4 + 7 x^2 + 8 x^4 + 4 x^6 + x^8)\denom$\\
			\hline
			$N= 8$ & $x^4 (10 + 17 x^2 + 18 x^4 + 8 x^6 + 3 x^8)\denom$\\
			\hline
			$N=6$ & $2 x^4 (4 + 7 x^2 + 8 x^4 + 4 x^6 + x^8)\denom+ {\underline {x^4(1+x^2+2x^4+x^6+x^8)\denom}}\,\,\,$\\
			\hline
			$N= 4$ & $3 x^4 (4 + 7 x^2 + 8 x^4 + 4 x^6 + x^8)\denom+ {\underline {x^6(2+2x^2+2x^4)\denom}}$\\
			\hline
		\end{tabular}\\
		\vspace{0.3cm}
		\begin{tabular}{|l|l|}
			\hline
			$SU(N)$ & Gluon partition function \\
			\hline
			$N\geq 4$ & $2 x^4 (4 + 7 x^2 + 8 x^4 + 4 x^6 + x^8)\denom+ {\underline {x^4(1+x^2+2x^4+x^6+x^8)\denom}}$ \\
			\hline
			$N=3$ & $x^4 (6 + 11 x^2 + 14 x^4 + 8 x^6 + x^8)\denom + {\underline {x^4(1+x^2+2x^4+x^6+x^8)\denom}}$\\
			\hline
			$N=2$ & $x^4 (4 + 7 x^2 + 8 x^4 + 4 x^6 + x^8)\denom$\\
			\hline
		\end{tabular}
	\end{center}
	\caption{Partition function over the space of Lagrangians in $D=6$ involving four $F^a_{\a\b}$'s. Recall $\denom\equiv 1/((1-x^4)(1-x^6))$. Contribution of $\CM^{\rm non-inv}$ is underlined. The rest is the contribution of $\CM^{\rm inv}$. Partition function for $SO(5)$ and $SO(7)$ is the same as that for $SO(N)$ for $N\geq 9$.}
	\label{gluonplethd6}
\end{table}

 \begin{table}
 	\begin{center}
 		\begin{tabular}{|l|l|}
 			\hline
 			$SO(N)$ & Gluon partition function   \\
 			\hline
 			$N\geq 9$ & $2 x^4 (4 + 7 x^2 + 7 x^4 + 3 x^6)\denom$\\
 			\hline
 			$N= 8$ & ${(2x^4(5+8x^2+8x^4+3x^6)+x^6)}\denom$\\
 			\hline
 			$N=6$ & $2 x^4 (4 + 7 x^2 + 7 x^4 + 3 x^6)\denom+ {\underline{x^5 (2 + 3 x^2 + x^3 + 3 x^4 + x^5 + x^6 + x^7)\denom}}$\\
 			\hline
 			$N= 4$ & $3 x^4 (4 + 7 x^2 + 7 x^4 + 3 x^6)\denom+ {\underline{x^5 (1 + x + 3 x^2 + x^3 + 3 x^4 + x^5 + 2 x^6)\denom}}$\\
 			\hline
 		\end{tabular}\\
 		\vspace{0.3cm}
 		\begin{tabular}{|l|l|}
 			\hline
 			$SU(N)$ & Gluon partition function  \\
 			\hline
 			$N\geq 4$ & $2 x^4 (4 + 7 x^2 + 7 x^4 + 3 x^6)\denom+ {\underline{x^5 (2 + 3 x^2 + x^3 + 3 x^4 + x^5 + x^6 + x^7)\denom}}\,$ \\
 			\hline
 			$N=3$ &$ x^4 (6 + 11 x^2 + 12 x^4 + 6 x^6)+ {\underline{x^5 (2 + 3 x^2 + x^3 + 3 x^4 + x^5 + x^6 + x^7)\denom}}$ \\
 			\hline
 			$N=2$ &$x^4 (4 + 7 x^2 + 7 x^4 + 3 x^6)\denom$ \\
 			\hline
 		\end{tabular}
 	\end{center}
 	\caption{Partition function over the space of Lagrangians in $D=5$ involving four $F^a_{\a\b}$'s. Recall $\denom\equiv 1/((1-x^4)(1-x^6))$. Contribution of $\CM^{\rm non-inv}$ is underlined. The rest is the contribution of $\CM^{\rm inv}$. Partition function for $SO(5)$ and $SO(7)$ is the same as that for $SO(N)$ for $N\geq 9$.}
 	\label{gluonplethd5}
 \end{table}

\begin{table}
	\begin{center}
		\begin{tabular}{|l|l|}
			\hline
			$SO(N)$ & Gluon partition function   \\
			\hline
			$N\geq 9$ & $2 x^4 (6 + 9 x^2 + 7 x^4 + x^6 - 2 x^8)\denom$\\
			\hline
			$N= 8$ & $x^4 (15 + 23 x^2 + 15 x^4 - 4 x^8)\denom$\\
			\hline
			$N=6$ & $2 x^4 (6 + 9 x^2 + 7 x^4 + x^6 - 2 x^8)\denom + {\underline{(x^2 + x^4 + x^6) (x^4 + 2 x^6)\denom}}\quad$\\
			\hline
			$N= 4$ &$3 x^4 (6 + 9 x^2 + 7 x^4 + x^6 - 2 x^8)\denom +  {\underline{(1+x^2 + x^4 ) (x^4 + 2 x^6)\denom}}$\\
			\hline
		\end{tabular}\\
		\vspace{0.3cm}
		\begin{tabular}{|l|l|}
			\hline
			$SU(N)$ & Gluon partition function \\
			\hline
			$N\geq 4$ & $2 x^4 (6 + 9 x^2 + 7 x^4 + x^6 - 2 x^8)\denom + {\underline{(x^2 + x^4 + x^6) (x^4 + 2 x^6)\denom}}$ \\
			\hline
			$N=3$ &$ x^4 (9 + 13 x^2 + 13 x^4 + 4 x^6 - 4 x^8)\denom + {\underline{(x^2 + x^4 + x^6) (x^4 + 2 x^6)\denom}}$ \\
			\hline
			$N=2$ &$  x^4 (6 + 9 x^2 + 7 x^4 + x^6 - 2 x^8)\denom$ \\
			\hline
		\end{tabular}
	\end{center}
	\caption{Partition function over the space of Lagrangians in $D=4$ involving four $F^a_{\a\b}$'s. Recall $\denom\equiv 1/((1-x^4)(1-x^6))$. Contribution of $\CM^{\rm non-inv}$ is underlined. The rest is the contribution of $\CM^{\rm inv}$. Partition function for $SO(5)$ and $SO(7)$ is the same as that for $SO(N)$ for $N\geq 9$.}
	\label{gluonplethd4}
\end{table}

\section{Explicit description of the gluon module}\label{explicit}
The gluon module $\CM^{\rm gluon}$ is the direct sum of $\CM^{\rm inv}$ and $\CM^{\rm non-inv}$. The module $\CM^{\rm inv}$ is the direct product of the photon module $\CM^{\rm photon}$ with the space of color structures $\CV^{\rm scalar}$ while module $\CM^{\rm non-inv}$ is constructed by taking the tensor product of non-quasi-invariant objects.

\subsection{$\CM^{\rm inv}$}\label{M-inv}
Let us first focus on $\CM^{\rm inv}=\CV^{\rm scalar}\otimes \CM^{\rm photon}$. We will describe the pieces of the tensor product explicitly. 

\subsubsection{Quasi-invariant scalar color structures i.e. basis of ${\CV^{\rm scalar}}$}\label{quasi-inv-color}
In this subsection we will list all the quasi-invariant color structures for $SO(N)$ and $SU(N)$. Their number is matched against the results of the table \ref{quasi-inv-counting} as well as of the table \ref{scalar-plethystic}. 
\subsubsection*{$\underline{SO(N)}$}
\subsubsection*{\underline{$N\geq 9$}}
For $N\geq 9$, there are two quasi-invariant color structures $\chi_{{\bf 3},1}$ and $\chi_{{\bf 3},2}$ both transforming under ${\bf 3}$.
\bea\label{simple-color}
\chi_{{\bf 3},1}^{(1)}&=&{\rm Tr}(\Phi_1\Phi_2){\rm Tr}(\Phi_3\Phi_4)\nonumber\\
\chi_{{\bf 3},2}^{(1)}&=&{\rm Tr}(\Phi_1\Phi_2 \Phi_3\Phi_4).
\eea
Both the structures are automatically symmetric under $\Z_2\times \Z_2$.
\subsubsection*{\underline{$N=8$}}
For $N=8$ there is an additional structure that transforms under ${\bf 1}_S$.
\be\label{qiso8}
\chi_{\bf S}^{SO(8)}=\Phi_1\wedge \Phi_2\wedge\Phi_3 \wedge \Phi_4.
\ee
where $\wedge$ is taken in the space of $SO(8)$ vector indices. This structure is automatically symmetric under $\Z_2\times \Z_2$.
\subsubsection*{\underline{$N=7,6,5$}}
For these values of $N$, the quasi-invariant color structures are the same as for $N\geq 9$.
\subsubsection*{\underline{$N=4$}}
For $SO(4)$ there is an additional color structure $\chi_{\bf 3}^{SO(4)}$ that transforms in ${\bf 3}$ compared to the structures in $N\geq 9$.
\be\label{qiso4sc}
\chi_{\bf 3}^{SO(4),(1)}=\Phi_1\wedge \Phi_2 {\rm Tr}(\Phi_3\Phi_4)|_{\Z_2\times \Z_2}.
\ee
Note that this generator is not automatically $\Z_2\times \Z_2$ symmetric and requires explicit symmetrization.

The counting of the quasi-invariant structures for $G=SO(N)$ indeed matches with the counting in tables \ref{quasi-inv-counting} and \ref{scalar-plethystic}.

\subsubsection*{$\underline{SU(N)}$}
\subsubsection*{\underline{$N\geq 4$}}
For $N\geq 4$, there are two quasi-invariant color structures $\xi_{{\bf 3},1}$ and $\xi_{{\bf 3},2}$ both transforming under ${\bf 3}$.
\bea\label{simple-color sun}
\xi_{{\bf 3},1}^{(1)}&=&{\rm Tr}(\Phi_1\Phi_2){\rm Tr}(\Phi_3\Phi_4)\nonumber\\
\xi_{{\bf 3},2}^{(1)}&=&{\rm Tr}(\Phi_1\Phi_2 \Phi_3\Phi_4)|_{\Z_2\times \Z_2}.
\eea
Here the first structure is automatically symmetric under $\Z_2\times \Z_2$ while the second one requires explicit symmetrization.
\subsubsection*{\underline{$N=3$}}
Compared to $N\geq 4$, the case of $N=3$ has one less generator. This is due to the special Jacobi relation of $SU(3)$,
\begin{eqnarray}\label{su3traceid}
d_{abf}d_{cdf}+ d_{acf}d_{bdf} +d_{adf}d_{bcf}= \frac{1}{3}\left(\delta_{ab}\delta_{cd}+\delta_{ac}\delta_{bd}+\delta_{ad}\delta_{bc})\right)
\end{eqnarray} 
This relates the ${\bf 1_S}$ parts of $\xi_{{\bf 3},1}$ and $\xi_{{\bf 3},2}$.
\subsubsection*{\underline{$N=2$}}
For $SU(2)$ the structures $\xi_{{\bf 3},1}$ and $\xi_{{\bf 3},2}$ are the same. This means there is only one quasi-invariant structure, say  $\xi_{{\bf 3},1}$. This is automatically $\Z_2\times \Z_2$ invariant.

The counting of the quasi-invariant structures for $G=SU(N)$ indeed matches with the counting in tables  \ref{quasi-inv-counting} and \ref{scalar-plethystic}.

\subsubsection{Photon module}\label{photon-module}
In this subsection we will summarize the results of \cite{Chowdhury:2019kaq} about the photon module in various dimensions. The photon module $\CM^{\rm photon}$ is any dimension is a direct sum of parity even module and parity odd module. The parity even module is generated freely for $D\geq 5$ by the same set of generators. For $D=4$, it gets relations. The parity odd module appears for $D\leq 7$ and depends crucially on dimension. We will first describe the parity even module and then the parity odd module.

\subsubsection*{\underline{Parity even}}
\subsubsection*{\underline{$D\geq 5$}}
In $D\geq 5$, the photon module is generated freely by the generators $E_{{\bf 3},1}, E_{{\bf 3},2}$ which transform in ${\bf 3}$ and $E_{\bf S}$ which transforms in ${\bf 1_S}$.\footnote{Here, $F_i^{\mu \nu}$ should be taken to mean $k_i^\mu \epsilon_i^\nu - k_i^\nu \epsilon_i^\mu$ and trace is take in the space of Lorentz indices.}
\bea\label{photon-lag}
E^{(1)}_{{\bf 3}, 1} &=& 8{\rm Tr}(F_1F_2){\rm Tr}(F_3F_4),~~~E^{(2)}_{{\bf 3}, 1} = 8{\rm Tr}(F_1F_3){\rm Tr}(F_2F_4),~~~E^{(3)}_{{\bf 3}, 1} =8{\rm Tr}(F_1F_4){\rm Tr}(F_3F_2),\\
E^{(1)}_{{\bf 3}, 2} &=& 8{\rm Tr}(F_1F_3F_2F_4),~~~~~~~~E^{(2)}_{{\bf 3}, 2} = 8{\rm Tr}(F_1F_2F_3F_4),~~~~~~~~E^{(3)}_{{\bf 3}, 2}= 8{\rm Tr}(F_1F_3 F_4F_2),\nonumber\\
E_{\bf S}&\simeq& -6F^{ab}_{1}  \partial_a F_{2}^{\mu\nu} \partial_b F_{3}^{\nu\rho} F_{4}^{\rho \mu}|_{\Z_2\times \Z_2}\nonumber\\
&=& 6\left(-F^{ab}_{1}  \partial_a F_{2}^{\mu\nu} \partial_b F_{3}^{\nu\rho} F_{4}^{\rho \mu} -F^{ab}_{2}  \partial_a F_{1}^{\mu\nu} \partial_b F_{4}^{\nu\rho} F_{3}^{\rho \mu}-F^{ab}_{3}  \partial_a F_{4}^{\mu\nu} \partial_b F_{1}^{\nu\rho} F_{2}^{\rho \mu}-F^{ab}_{4}  \partial_a F_{3}^{\mu\nu} \partial_b F_{2}^{\nu\rho} F_{1}^{\rho \mu}\right). \nonumber
\eea
The superscript $(i)$ label the three states of the representation ${\bf 3}$. Note that every state in $E_{{\bf 3},1}$ and $E_{{\bf 3},2}$ is automatically $\Z_2\times \Z_2$ invariant while the state in $E_{\bf S}$ requires explicit $\Z_2\times\Z_2$ symmetrization.

\subsubsection*{\underline{$D=4$}}
In $D=4$ these generators have relations,
\be\label{4d-relation}
\big(\sum_{i=1,2,3}~ \frac12 \cf^{E_{{\bf 3},1}^{(i)}}(t,u) E_{{\bf 3},1}^{(i)}\big)+\big(\sum_{i=1,2,3} ~ \frac12\cf^{E_{{\bf 3},2}^{(i)}}(t,u) E_{{\bf 3},2}^{(i)}\big)=0.
\ee
where the functions $\cf$ are, 
\bea \label{defouramb}
\cf_1^{E_{{\bf 3},1}}(t,u)&=&\frac18(s^2+t^2+u^2)f(t,u),\nonumber\\  
\cf_1^{E_{{\bf 3},2}}(t,u)&=& t u f(t,u).
\eea
and 
\bea
\cf_2^{E_{{\bf 3},1}}(t,u)&=&\frac98 s tu \,g(t,u),\nonumber\\
\cf_2^{E_{{\bf 3},2}}(t,u)&=&-\frac12(2(t^3+u^3)-stu) \,g(t,u).
\eea
where $f(t,u)$ and $g(t,u)$ are arbitrary functions that are symmetric in the two arguments. These two relations transform in ${\bf 1_S}$.

\subsubsection*{\underline{Parity odd}}
The parity odd module is present only for $D\leq 7$. Let us start with $D=7$ and work our way down.
\subsubsection*{\underline{$D=7$}}
In $D=7$ the parity odd module is generated freely by a single generator $O_{\bf S}^{D=7}$ transforming in ${\bf 1_S}$.
\be
O_{\bf S}^{D=7}=*(A_1\wedge F_2\wedge F_3\wedge F_4).
\ee
It is automatically $\Z_2\times \Z_2$ invariant.
\subsubsection*{\underline{$D=6$}}
In $D=6$ the parity odd module is generated freely by a single generator $O_{\bf A}^{D=6}$ transforming in ${\bf 1_A}$.
\be\label{D6odd}
O_{\bf A}^{D=6}=F^{ab}_1*(\partial_a F_2\wedge \partial_b F_3\wedge F_4)|_{\Z_2\times \Z_2}.
\ee
The resulting S-matrix is not automatically $\Z_2\times \Z_2$ symmetric but needs a projection onto $\Z_2\times \Z_2$ symmetric state.
\subsubsection*{\underline{$D=5$}}
There is no parity odd quasi-invariant module for $D=5$.
\subsubsection*{\underline{$D=4$}}
The parity odd module for $D=4$ is the most complicated. It is generated by $O_{\bf 3}^{D=4}$ which transforms in ${\bf 3}$ and $O_{\bf S}^{D=4}$ which transforms in ${\bf 1_S}$.
\be\label{podfgen} 
\begin{split}
	O_{\bf 3}^{D=4,(1)}&\equiv 2*(F_1\wedge F_2){\rm Tr}(F_3 F_4)|_{\Z_2\times \Z_2}=4*(F_1\wedge F_2){\rm Tr}(F_3 F_4)+4*(F_3\wedge F_4){\rm Tr}(F_1 F_2),\nonumber\\
	O_{\bf S}^{D=4}&\equiv 6\varepsilon_{\mu\nu\rho\sigma}F_1^{\mu\nu} \partial^\rho F_2^{ab}\partial^\sigma F_3^{bc} F_4^{ca}|_{\Z_2\times \Z_2}=6\bigg(\varepsilon_{\mu\nu\rho\sigma}F_1^{\mu\nu} \partial^\rho F_2^{ab}\partial^\sigma F_3^{bc} F_4^{ca}+ \varepsilon_{\mu\nu\rho\sigma}F_2^{\mu\nu} \partial^\rho F_1^{ab}\partial^\sigma F_4^{bc} F_3^{ca}\nonumber\\
	&+ \varepsilon_{\mu\nu\rho\sigma}F_3^{\mu\nu} \partial^\rho F_4^{ab}\partial^\sigma F_1^{bc} F_2^{ca}+\varepsilon_{\mu\nu\rho\sigma}F_4^{\mu\nu} \partial^\rho F_3^{ab}\partial^\sigma F_2^{bc} F_1^{ca}\bigg).
\end{split}
\ee
Note that neither of these generators is automatically $\Z_2\times \Z_2$ invariant and requires a projection. The generator $O_{\bf 3}^{D=4}$ enjoys a relation.
\be\label{poin}
\sum_{i=1,2,3} ~\frac12\cf^{O_{\bf 3}^{(i)}}(t,u)O_{\bf 3}^{(i)}=0.
\ee
where the functions $\cf$ are 
\be \label{solpoin} 
\cf_1^{O_{\bf 3}}(t,u) =2 t u f(t,u) \quad{\rm or}\quad \cf_2^{O_{\bf 3}}(t,u) = (2(t^3+u^3)+(t+u)tu)g(t,u).
\ee
where $f(t,u)$ and $g(t,u)$ are arbitrary functions that are symmetric in the two arguments. Both these relations transform in ${\bf 1_S}$.

In summary, we tabulate the $S_3$ representation and the derivative order of generators as well as relations in table \ref{photon-table}.
\begin{table}
	\begin{center}
		\begin{tabular}{|l|l|l|l|}
			\hline
			Parity even & $n_{\bf S}$ & $n_{\bf M}$ & $n_{\bf A}$ \\
			\hline
			$D\geq 5$ & $2_{4\partial}+1_{6\partial}$ & $2_{4\partial}$ &0\\
			\hline
			$D=4$ & $2_{4\partial}+1_{6\partial}-1_{8\partial} -1_{10\partial}$ &  $2_{4\partial}$ & 0\\
			\hline
		\end{tabular}\\
		\vspace{0.3cm}
		\begin{tabular}{|l|l|l|l|}
			\hline
			Parity odd & $n_{\bf S}$ & $n_{\bf M}$ & $n_{\bf A}$ \\
			\hline
			$D\geq 8$ & 0& 0 &0\\
			\hline
			$D=7$ & $1_{3\partial}$ & 0& 0\\
			\hline
			$D=6$ & 0 & 0 & $1_{6\partial}$ \\
			\hline
			$D=5$ & 0 & 0 & 0 \\
			\hline
			$D=4$ & $1_{4\partial}+1_{6\partial}-1_{8\partial} -1_{10\partial}\,$ & $1_{4\partial}$ & 0 \\
			\hline
		\end{tabular}
	\end{center}
	\caption{The counting of generators and relations for the quasi-invariant local photon module. The relations are counted with negative sign. The subscript $\partial$ indicates the derivative order of the generators/relations.}
	\label{photon-table}
\end{table}

\subsubsection{Tensor product}
Now that we understand the photon module (detailed in section \ref{photon-module}) and the quasi-invariant structure (detailed in section \ref{quasi-inv-color}), we can straightforwardly construct $\CM^{\rm inv}$ which is a part of the gluon module by taking their tensor product. We will not carry this out explicitly in all dimensions but rather illustrate the procedure for the case of $D\geq 8$ and for $G=SO(N)$ and $G=SU(N)$.

For $SO(N)$ for $N\geq 9$, the photon module is freely generated by, $E_{{\bf 3},1}, E_{{\bf 3},2}$ at $4$ derivatives and $E_{\bf S}$ at $6$ derivatives, given in equation \eqref{photon-lag}. The color structures are $\chi_{{\bf 3},1}$ and $\chi_{{\bf 3},2}$, given in equation \eqref{simple-color}. To compute their tensor product, we would need to understand the decomposition of ${\bf 3}\otimes {\bf 3}$. Instead of decomposing this tensor product in irreducible representation, it is more intuitive to understand it as ${\bf 3}\otimes {\bf 3}=2\cdot {\bf 3}+{\bf 3_A}$ (using Clebsch-Gordan rules \eqref{cgoeff} and \eqref{cgoeff1}). The states on the right-hand side are constructed explicitly as follows. 

Consider the tensor product of $|e^{(i)}\rangle \in {\bf 3}$ and $|f^{(i)}\rangle \in {\bf 3}$. The three  representations on the right hand side of the product are
\bea
&&|e^{(1)}f^{(1)}\rangle, \ldots \nonumber\\
&& |e^{(1)}(f^{(2)}+f^{(3)})\rangle, \ldots \nonumber\\
&& |e^{(1)}(f^{(2)}-f^{(3)})\rangle, \ldots
\eea
The first two representations transform as ${\bf 3}$ and the last one is a ${\bf 3_A}$.

Getting back to the problem at hand, we need to compute the tensor product,  $({\bf 3}\oplus{\bf 3})_{\rm photon}\otimes ({\bf 3}\oplus {\bf 3})_{\rm scalar}$ at $4$ derivatives and  $({\bf 1_S})_{\rm photon}\otimes ({\bf 3}\oplus {\bf 3})_{\rm scalar}$ at $6$ derivatives. The second tensor product is trivial $({\bf 1_S})_{\rm photon}\otimes ({\bf 3}\oplus {\bf 3})_{\rm scalar}=2\cdot {\bf 3}$. The first one is computed using the above conventions. We get  $({\bf 3}\oplus{\bf 3})_{\rm photon}\otimes ({\bf 3}\oplus {\bf 3})_{\rm scalar}=8 \cdot {\bf 3}+4\cdot {\bf 3_A}$. All in all, we get $10\cdot {\bf 3}\oplus 4\cdot {\bf 3_A}$.

Explicitly, for $SO(N)$ for $N\geq 9$, the following Lagrangians give rise to these generators,

\begingroup
\allowdisplaybreaks
\begin{eqnarray}\label{plagsonadj}
	G^1_{SO(N)} &=& \sum_{m,n}  a_{m,n} \prod_{b=1}^m\prod_{c=1}^n {\rm Tr} \left( T^e T^f\right) {\rm Tr} \left( T^g T^h\right) ~\left(\partial_{\m_b} \partial_{\n_c} F^e_{ij} F^f_{ji} \right)  \left(\partial^{\m_b} F^g_{kl} \partial^{\n_c} F^h_{lk} \right) \nonumber\\
	G^2_{SO(N)} &=& \sum_{m,n}  a_{m,n} \prod_{b=1}^m\prod_{c=1}^n \left({\rm Tr} \left( T^e T^g\right) {\rm Tr} \left( T^h T^f\right) +{\rm Tr} \left( T^e T^h\right) {\rm Tr} \left( T^f T^g\right) \right)\left(\partial_{\m_b} \partial_{\n_c} F^e_{ij} F^f_{ji} \right)  \left(\partial^{\m_b} F^g_{kl} \partial^{\n_c} F^h_{lk} \right) \nonumber\\
	G^3_{SO(N)} &=& \sum_{m,n}  a_{m,n} \prod_{b=1}^m\prod_{c=1}^n \left({\rm Tr} \left( T^e T^g\right) {\rm Tr} \left( T^h T^f\right) -{\rm Tr} \left( T^e T^h\right) {\rm Tr} \left( T^f T^g\right) \right)~\left(\partial_{\m_b} \partial_{\n_c} F^e_{ij} F^f_{ji} \right)  \left(\partial^{\m_b} F^g_{kl} \partial^{\n_c} F^h_{lk} \right) \nonumber\\
	G^4_{SO(N)} &=& \sum_{m,n}  a_{m,n} \prod_{b=1}^m\prod_{c=1}^n \left({\rm Tr} \left( T^e T^g T^f T^h \right)+{\rm Tr} \left( T^e T^h T^f T^g \right) \right) ~\left(\partial_{\m_b} \partial_{\n_c} F^e_{ij} F^f_{ji} \right)  \left(\partial^{\m_b} F^g_{kl} \partial^{\n_c} F^h_{lk} \right) \nonumber\\
	G^5_{SO(N)} &=& \sum_{m,n}  a_{m,n} \prod_{b=1}^m\prod_{c=1}^n {\rm Tr}(\{T^e, T^f \}\{T^g, T^h\})~\left(\partial_{\m_b} \partial_{\n_c} F^e_{ij} F^f_{ji} \right)  \left(\partial^{\m_b} F^g_{kl} \partial^{\n_c} F^h_{lk} \right) \nonumber\\
	G^6_{SO(N)} &=& \sum_{m,n}  a_{m,n} \prod_{b=1}^m\prod_{c=1}^n{\rm Tr}(\left[T^e, T^f \right]\left[T^g, T^h\right])\left(\partial_{\m_b} \partial_{\n_c} F^e_{ij} F^f_{ji} \right)  \left(\partial^{\m_b} F^g_{kl} \partial^{\n_c} F^h_{lk} \right) \nonumber\\
	G^7_{SO(N)} &=& \sum_{m,n}  a_{m,n} \prod_{b=1}^m\prod_{c=1}^n {\rm Tr} \left( T^e T^f\right) {\rm Tr} \left( T^g T^h\right) ~\left(\partial_{\m_b} \partial_{\n_c} F^e_{ij} \partial^{\m_b}F^g_{jk} F^f_{kl} \partial^{\n_c} F^h_{li} \right) \nonumber\\
	G^8_{SO(N)} &=& \sum_{m,n}  a_{m,n} \prod_{b=1}^m\prod_{c=1}^n\left( {\rm Tr} \left( T^e T^g\right) {\rm Tr} \left( T^h T^f\right) + {\rm Tr} \left( T^e T^h\right) {\rm Tr} \left( T^f T^g\right) \right)\left(\partial_{\m_b} \partial_{\n_c} F^e_{ij} \partial^{\m_b}F^f_{jk} F^g_{kl} \partial^{\n_c} F^h_{li} \right) \nonumber\\
	G^9_{SO(N)} &=& \sum_{m,n}  a_{m,n} \prod_{b=1}^m\prod_{c=1}^n \left( {\rm Tr} \left( T^e T^g\right) {\rm Tr} \left( T^h T^f\right) - {\rm Tr} \left( T^e T^h\right) {\rm Tr} \left( T^f T^g\right) \right)\left(\partial_{\m_b} \partial_{\n_c} F^e_{ij} \partial^{\m_b}F^f_{jk} F^g_{kl} \partial^{\n_c} F^h_{li} \right) \nonumber\\
	G^{10}_{SO(N)} &=& \sum_{m,n}  a_{m,n} \prod_{b=1}^m\prod_{c=1}^n \left({\rm Tr} \left( T^e T^f T^g T^h \right)+{\rm Tr} \left( T^e T^h T^g T^f \right) \right) ~\left(\partial_{\m_b} \partial_{\n_c} F^e_{ij} \partial^{\m_b}F^f_{jk} F^g_{kl} \partial^{\n_c} F^h_{li} \right)\nonumber\\
	G^{11}_{SO(N)} &=& \sum_{m,n}  a_{m,n} \prod_{b=1}^m\prod_{c=1}^n{\rm Tr}( \{T^e, T^g \}\{T^h,T^f\})~\left(\partial_{\m_b} \partial_{\n_c} F^e_{ij} \partial^{\m_b}F^f_{jk} F^g_{kl} \partial^{\n_c} F^h_{li} \right) \nonumber\\
	G^{12}_{SO(N)} &=& \sum_{m,n}  a_{m,n} \prod_{b=1}^m\prod_{c=1}^n  {\rm Tr}(\left[T^e, T^g \right]\left[T^h, T^f\right])
	~\left(\partial_{\m_b} \partial_{\n_c} F^e_{ij} \partial^{\m_b}F^f_{jk} F^g_{kl} \partial^{\n_c} F^h_{li} \right) \nonumber\\
	G^{13}_{SO(N)} &=& \sum_{m,n}  a_{m,n} \prod_{b=1}^m\prod_{c=1}^n {\rm Tr} \left( T^e T^f\right) {\rm Tr}\left( T^g T^h\right) ~\left(\partial_{\m_b} \partial_{\n_c} F^e_{\a\b} \partial_{\a}F^f_{jk} \partial^{\m_b}\partial_{\b}F^g_{kl} \partial^{\n_c} F^h_{lj} \right) \nonumber\\
	G^{14}_{SO(N)} &=& \sum_{m,n}  a_{m,n} \prod_{b=1}^m\prod_{c=1}^n {\rm Tr}(\{T^e, T^f \}\{T^g, T^h\} )~\left(\partial_{\m_b} \partial_{\n_c} F^e_{\a\b} \partial_{\a}F^f_{jk} \partial^{\m_b}\partial_{\b}F^g_{kl} \partial^{\n_c} F^h_{lj} \right).
\end{eqnarray}
\endgroup
Here $T^a$ are the $SO(N)$ generators and we have used the following condensed notation for the derivatives,
\be
\prod_{b=1}^m \,\,\partial_{\mu_b} \,{\cal O}_1\, \partial^{\mu_b} \,{\cal O}_2 \equiv \partial_{\mu_1}\partial_{\mu_2}\ldots \partial_{\mu_m} \,{\cal O}_1\,\,\partial^{\mu_1}\partial^{\mu_2}\ldots \partial^{\mu_m} \,{\cal O}_2.
\ee
for some operators ${\cal O}_1$ and ${\cal O}_2$. The same notation is also used for the second tower of derivatives indexed as $\partial_{\nu_c}$. In particular, each term denotes a Lorentz invariant Lagrangian term with $2m+2n$ derivatives. As these terms are supposed to represent S-matrix we have only required them to be invariant under linearized gauge transformations but if one wants associated Lagrangians that are invariant full non-linear gauge transformations, one simple replaces product of ordinary derivatives with symmetrized product of covariant derivatives.\footnote{Once made invariant under nonlinear gauge transformations, the Lagrangians in equation \eqref{plagsonadj} also give rise to higher point S-matrices. Note that this list will not be exhaustive because the covariant derivatives have been symmetrized and hence the Lagrangians always contain only four Field strength operators $F_{\mu\nu}$. However if one relaxes the condition of symmetrization, one can generate all the Lagrangians as the field strength can be constructed as $[D_\mu, D_\nu]$.} This discussion also applies to the Lagrangian terms listed in Appendix \ref{explicit-gluon-lag}.

Among these Lagrangians in equation \eqref{plagsonadj}, $G_{SO(N)}^{3,6,9,12}$ give rise to quasi-invariant generators transforming in ${\bf 3_A}$ and the rest give rise to ${\bf 3}$.
The results for $\CM^{\rm inv}$ generators for $SO(N<9)$ and $D\geq8$ are presented in appendix \ref{egmsnso}.

For $SU(N)$ ($N\geq 4$), the color structures are $\xi_{{\bf 3},1}$ and $\xi_{{\bf 3},2}$, given in equation \eqref{simple-color sun}. In their tensor product, we get $({\bf 3}\oplus{\bf 3})_{\rm photon}\otimes ({\bf 3}\oplus {\bf 3})_{\rm scalar}=8 \cdot {\bf 3}+4\cdot {\bf 3_A}$ generators at 4 derivatives and $({\bf 1_S})_{\rm photon}\otimes ({\bf 3}\oplus {\bf 3})_{\rm scalar}=2\cdot {\bf 3}$ generators at $6$ derivatives.  This counting is the same as that for the case of $SO(N\geq 9)$. The explicit Lagrangians that give rise to these generators are also the same ones as those for $SO(N\geq 9)$ given in equation \eqref{plagsonadj}. We simply need to interpret $T^a$'s as generators of $SU(N)$. We present the explicit $\CM^{\rm inv}$ for $SU(N<4)$,  $D\geq8$ gluon modules in appendix \ref{egmsnsu}.

We encode the $S_3$ representation and the derivative order of the gluon generators by specifying the triplet $(n_{\bf S},n_{\bf M},n_{\bf A})=(8_{4\partial}+2_{6\partial},12_{4\partial}+2_{6\partial},4_{4\partial})$. The counting of gluon generators of $\CM^{\rm inv}$ for other gauge groups is along with their $S_3$ representation and the derivative order are tabulated in table \ref{M-inv-table}. For simplicity, we have only done this for $D\geq 8$. Similar tabulation can also be done for $D<8$ by taking the tensor product of appropriate generators in table \ref{photon-table} with the quasi-invariant color structures in table \ref{quasi-inv-counting}. 
\begin{table}
	\begin{center}
		\begin{tabular}{|l|l|l|l|}
			\hline
			$SO(N)$ & $n_{\bf S}$ & $n_{\bf M}$ & $n_{\bf A}$ \\
			\hline
			$N\geq  9$ & $8_{4\partial}+2_{6\partial}$& $12_{4\partial}+2_{6\partial}$ &$4_{4\partial}$\\
			\hline
			$N=8$ & $10_{4\partial}+3_{6\partial}$ & $14_{4\partial}+2_{6\partial}$ & $4_{4\partial}$\\
			\hline
			$N=4$ & $12_{4\partial}+3_{6\partial}$ & $18_{4\partial}+3_{6\partial}$ & $6_{4\partial}$\\
			\hline
		\end{tabular}\\
		\vspace{0.3cm}
		\begin{tabular}{|l|l|l|l|}
			\hline
			$SU(N)$ & $n_{\bf S}$ & $n_{\bf M}$ & $n_{\bf A}$ \\
			\hline
			$N\geq  4$ & $8_{4\partial}+2_{6\partial}$\,\,& $12_{4\partial}+2_{6\partial}\,$ &$4_{4\partial}$\\
			\hline
			$N=3$ & $6_{4\partial}+1_{6\partial}$ & $10_{4\partial}+2_{6\partial}$ & $4_{4\partial}$\\
			\hline
			$N=2$ & $4_{4\partial}+1_{6\partial}$ & $6_{4\partial}+1_{6\partial}$ & $2_{4\partial}$\\
			\hline
		\end{tabular}
	\end{center}
	\caption{The number of generators of $\CM^{\rm inv}$ in $D\geq 8$. We have indicated the $S_3$ representation and the derivative order of all the generators. The counting for $SO(N)$ with $N=7,6,5$ is the same as that for $N\geq 9$.}
	\label{M-inv-table}
\end{table}
Alternatively, one can encode the $S_3$ representations and the derivatives of the $\CM^{\rm inv}$ generators into a partition function. We have done that for all cases, including $D<8$, and have identified their contribution to tables  \ref{gluon-plethystic}, \ref{gluonplethd7}, \ref{gluonplethd6}, \ref{gluonplethd5} and \ref{gluonplethd4}. The remaining contribution is underlined and is attributed to $\CM^{\rm non-inv}$. 

Now we will construct generators of $\CM^{\rm non-inv}$. We will confirm its construction by matching the partition function over it with the underlined part of the tables  \ref{gluon-plethystic}, \ref{gluonplethd7}, \ref{gluonplethd6}, \ref{gluonplethd5} and \ref{gluonplethd4}.

\subsection{$\CM^{\rm non-inv}$}
This part of the gluon module comes from the tensor product of non-quasi-invariant generators of photons and non-quasi-invariant color structures. In section \ref{M-inv}, when we explicitly listed the quasi-invariant generators and quasi-invariant color structures, we kept track of whether the generator/color-structure is automatically $\Z_2\times \Z_2$ invariant or it needs to be $\Z_2\times \Z_2$ symmetrized by hand. The non-quasi-invariant structures are obtained from the later type by projecting onto states with charge, say $(+--)$ under $\Z_2\times \Z_2$ rather than onto states with charge $(+++)$. Moreover, for the state with charge $(+--)$ we will also look at the charge under $P_{12}$ to deduce whether the state belongs to ${\bf 3}$ or ${\bf 3_A}$ of $S_3$. However, there may be non-quasi-invariant states that can not be constructed using the above strategy. The analysis of such states is done case by case below.
As in section \ref{M-inv} we will first focus on scalars and then on photons.

\subsubsection{Non-quasi-invariant color structures}\label{non-inv-color}
In this subsection we will list all the quasi-invariant color structures for $SO(N)$ and $SU(N)$. Their number is matched against the results of the tables \ref{non-quasi-inv-counting}. 
\subsubsection*{$\underline{SO(N)}$}
\subsubsection*{\underline{$N\geq 7$}}
For $N\geq 7$, we do not find any non-quasi-invariant structures.
\subsubsection*{\underline{$N=6$}}
We do find a non-quasi-invariant color structure for $SO(6)$. It is convenient to describe it using $SO(6)=SU(4)$ language. We will do so shortly.
\subsubsection*{\underline{$N=4$}}
For $SO(4)$, the additional color structure $\chi_{\bf 3}^{SO(4)}$ is not automatically $\Z_2\times \Z_2$ symmetric and requires explicit symmetrization. Projecting on the state with charge $(+--)$ instead,
\bea\label{qniso4sc}
{\tilde \chi}^{SO(4),(1)}=\Phi_1\wedge \Phi_2 {\rm Tr}(\Phi_3\Phi_4)|_{(+--)}&=&\Phi_1\wedge \Phi_2 {\rm Tr}(\Phi_3\Phi_4)+\Phi_2\wedge \Phi_1 {\rm Tr}(\Phi_4\Phi_3)\nonumber\\
&-&\Phi_3\wedge \Phi_4 {\rm Tr}(\Phi_1\Phi_2)-\Phi_4\wedge \Phi_3 {\rm Tr}(\Phi_2\Phi_1).
\eea
This structure is symmetric under $P_{12}$ hence transforms as ${\bf 3}$. We denote it as ${\tilde \chi}_{\bf 3}^{SO(4)}$.

The counting of the non-quasi-invariant structures for $G=SO(N)$ indeed matches with the counting in table \ref{non-quasi-inv-counting}. 

\subsubsection*{$\underline{SU(N)}$}
\subsubsection*{\underline{$N\geq 3$}}
For $N\geq 3$, among the two quasi-invariant color structures, $\xi_{{\bf 3},2}$ requires explicit $\Z_2\times \Z_2$ symmetrization. Projecting onto the state with $(+--)$ charge instead,
\bea \label{mnisun}
{\tilde \xi}_{{\bf 3_A}}^{(1)}={\rm Tr}(\Phi_1\Phi_2 \Phi_3\Phi_4)|_{(+--)}&=&{\rm Tr}(\Phi_1\Phi_2 \Phi_3\Phi_4)+{\rm Tr}(\Phi_2\Phi_1 \Phi_4\Phi_3)\nonumber\\
&&- {\rm Tr}(\Phi_3\Phi_4 \Phi_1\Phi_2) - {\rm Tr}(\Phi_4\Phi_3 \Phi_2\Phi_1).
\eea
This state is anti-symmetric under $P_{12}$ hence transforms as ${\bf 3}_A$. We denote it as ${\tilde \xi}_{\bf 3_A}$.
\subsubsection*{\underline{$N=2$}}
The only quasi-invariant structure  $\xi_{{\bf 3},1}$ is automatically $\Z_2\times \Z_2$ invariant. Hence, there is no non-quasi-invariant structure.

The counting of the quasi-invariant structures for $G=SU(N)$ indeed matches with the counting in table \ref{non-quasi-inv-counting}.

\subsubsection{Non-quasi-invariant photon structures}\label{non-inv-photon}
In this section, we will construct the non-quasi-invariant photon structures first for the parity even case and then for the parity odd case. The strategy of constructing them from quasi-invariants will not always work. As we will see, there will be new parity odd non-quasi-invariant states in lower dimensions.

\subsubsection*{\underline{Parity even}}
\subsubsection*{\underline{$D\geq 4$}}
In $D\geq 5$, the parity even generators $E_{{\bf 3},1}, E_{{\bf 3},2}$ are automatically quasi-invariant but $E_{\bf S}$ requires explicit symmetrization. In order to get the non-quasi-invariant structure we will project the non-projected S-matrix of $E_{\bf S}$ onto state with $(+--) $ charge.
\begin{eqnarray}\label{photon-lag2}
{\tilde E}^{(1)}&\simeq& -6F^{ab}_{1}  \partial_a F_{2}^{\mu\nu} \partial_b F_{3}^{\nu\rho} F_{4}^{\rho \mu}|_{(+--)}\nonumber\\
&=& 6\left(-F^{ab}_{1}  \partial_a F_{2}^{\mu\nu} \partial_b F_{3}^{\nu\rho} F_{4}^{\rho \mu} -F^{ab}_{2}  \partial_a F_{1}^{\mu\nu} \partial_b F_{4}^{\nu\rho} F_{3}^{\rho \mu}+F^{ab}_{3}  \partial_a F_{4}^{\mu\nu} \partial_b F_{1}^{\nu\rho} F_{2}^{\rho \mu}+F^{ab}_{4}  \partial_a F_{3}^{\mu\nu} \partial_b F_{2}^{\nu\rho} F_{1}^{\rho \mu}\right). \nonumber\\
\end{eqnarray}
We observe that this structure is symmetric under $P_{12}$. This means $({\tilde E}^{(1)},{\tilde E}^{(2)},{\tilde E}^{(3)})$ form representation ${\bf 3}$ of $S_3$. We denote this as ${\tilde E}_{\bf 3}$. 


\subsubsection*{\underline{$D=4$}}
In $D=4$, the quasi-invariant generators $E_{{\bf 3},1}, E_{{\bf 3},2}$ obey a relation. The generator  $E_{\bf S}$ is not part of this relation, hence the term \eqref{photon-lag2} is a non-quasi-invariant transforming in ${\bf 3}$ even in $D=4$.
%

\subsubsection*{\underline{Parity odd}}
There is no parity odd  module in $D\geq 8$.
\subsubsection*{\underline{$D=7$}}
 There is a single parity odd generator in $D=7$, $O_{\bf S}^{D=7}$. 

\begin{equation} \label{cssevdef}
O_{\bf S}^{D=7}={\rm CS}_7= *\left( A_1\wedge F_2\wedge F_3 \wedge F_4 \right) 
\end{equation} 
It is automatically $\Z_2\times \Z_2$ invariant. Hence, while it contributes to the quasi-invariant structures, it does not contribute to the non-quasi-invariant structures. 

\subsubsection*{\underline{$D=6$} }


In $D=6$, we encountered a single quasi-invariant generator transforming in ${\bf 1}_A$,
\be
O^{D=6}_{\bf A}=F_1^{ab}*(\partial_a F_2\wedge \partial_b F_3 \wedge F_4)
\ee
This state requires explicit $\Z_2\times \Z_2$ symmetrization. Projecting onto state with $(+--)$ charge instead we get,
\bea
{{\tilde O}}^{'D=6,(1)}&=&F^{ab}_1*(\partial_a F_2\wedge \partial_b F_3\wedge F_4)+F^{ab}_2*(\partial_a F_1\wedge \partial_b F_4\wedge F_3)\nonumber\\
&-&F^{ab}_3*(\partial_a F_4\wedge \partial_b F_1\wedge F_2)-F^{ab}_4*(\partial_a F_3\wedge \partial_b F_2\wedge F_1)
\eea
This is a non-quasi-invariant S-matrix transforming as ${\bf 3_A}$. Interestingly, it turns out that this is not the most basic non-quasi-invariant S-matrix. We discuss this below.

Consider the S-matrix,
\be
\epsilon^{abcdef}F^{ab}_1F^{e\a}_2F^{\a f}_3F^{cd}_4.
\ee
It is not $\Z_2\times \Z_2$ symmetric. What is more is that it vanishes under $\Z_2\times \Z_2$ symmetrization so it does not contribute to the quasi-invariant module. However its projection onto $(+--)$ state is non-vanishing. We get,
\be
{{\tilde O}}^{D=6,(1)}=\epsilon^{abcdef}F^{ab}_1F^{e\a}_2F^{\a f}_3F^{cd}_4+\epsilon^{abcdef}F^{ab}_2F^{e\a}_1F^{\a f}_4F^{cd}_3.
\ee
It can be checked easily that ${{\tilde O}}^{'D=6,(1)}=(s/2){{\tilde O}}^{D=6,(1)}$. This means that the most basic non-quasi-invariant structure is ${{\tilde O}}^{D=6,(i)}$ and not ${{\tilde O}}^{'D=6,(i)}$. It transforms as $\bf 3_A$.

\subsubsection*{\underline{$D=5$}}

In $D=5$ there is no quasi-invariant parity odd module. However the non-quasi-invariant modules do exist as we spell out below.

Consider the S-matrices
\be\label{d5polag}
\epsilon^{abcde}F_{ab}^1F_{cd}^2\partial_eF_{\a\b}^3F_{\a\b}^4, \qquad \epsilon_{abcde}\partial_\b F_{ab}^1F_{cd}^2F_{e\a}^3 F_{\a\b}^4.
\ee
Both these vanish under $\Z_2\times \Z_2$ symmetrization but survive the projection onto $(+--)$ state. The $S_3$ orbit of the first S-matrix is $6$ dimensional and decomposes as ${\bf 3}\oplus (\bf 3_A)$ while the $S_3$ representation of the second S-matrix transforms in ${\bf 3}_A$. All in all there are ${\bf 3}\oplus 2\cdot {\bf 3_A}$ non-quasi-invariant structures in $D=5$, all of them at five derivatives. They are given below:
\bea
\tilde{{O}}^{D=5,(1)}_{{\bf 3}}&=& \left( \epsilon^{abcde}F^1_{ab}F^3_{cd} \partial^2_e F^2_{\a\b}F^4_{\a\b} + \epsilon^{abcde}F^1_{ab}F^4_{cd} \partial^2_e F^2_{\a\b}F^3_{\a\b} \right)|_{(+--)}, \nonumber\\
\tilde{{O}}^{D=5,(1)}_{{\bf 3_A},2}&=& \left( \epsilon^{abcde}F^1_{ab}F^3_{cd} \partial^2_e F^2_{\a\b}F^4_{\a\b} - \epsilon^{abcde}F^1_{ab}F^4_{cd} \partial^2_e F^2_{\a\b}F^3_{\a\b} \right)|_{(+--)}, \nonumber\\
\tilde{{O}}^{D=5,(1)}_{{\bf 3_A},1}&=& \left( \epsilon_{abcde}\partial^1_\b F^1_{ab}F^3_{cd}F^2_{e\a} F^4_{\a\b} + \epsilon_{abcde}\partial^2_\b F^2_{ab}F^4_{cd}F^1_{e\a} F^3_{\a\b} \right)|_{(+--)}.
\eea
Existence of these structures can be anticipated using the so called ``bare module" generators given in appendix D.1 of \cite{Chowdhury:2019kaq} as described in the footnote.\footnote{
The bare parity odd non-quasi-invariant S-matrices are
\be
\left( {\tilde \varepsilon}^2_{\mu\nu}
\polo_1\,^\mu \polo_3\,^\nu \a_2\a_4\right)|_{(+--)},\qquad \left( {\tilde \varepsilon}^2_{\mu\nu} 
\polo_1\,^\mu \polo_3\,^\nu \right) \left( \polo_2 . \polo_4 \right)|_{(+--)}.
\ee
The $S_3$ orbit of the first S-matrix is $6$ dimensional and decomposes as ${\bf 3}\oplus (\bf 3_A)$ while the $S_3$ representation of the second S-matrix transforms in ${\bf 3}_A$. 
Explicitly,
\bea\label{d5pobm1}
\tilde{o}^{D=5,(1)}_{{\bf 3}}&=& \left( {\tilde \varepsilon}^2_{\mu\nu}
\polo_1\,^\mu \polo_3\,^\nu \a_2\a_4 - {\tilde \varepsilon}^2_{\mu\nu}
\polo_1\,^\mu \polo_4\,^\nu \a_2\a_3 \right)|_{(+--)}, \nonumber\\
\tilde{o}^{D=5,(1)}_{{\bf 3_A},2}&=& \left( {\tilde \varepsilon}^2_{\mu\nu}
\polo_1\,^\mu \polo_3\,^\nu \a_2\a_4 + {\tilde \varepsilon}^2_{\mu\nu}
\polo_1\,^\mu \polo_4\,^\nu \a_2\a_3 \right)|_{(+--)}\nonumber\\
\tilde{o}^{D=5,(1)}_{{\bf 3_A},1}&=& \left( {\tilde \varepsilon}^2_{\mu\nu} 
\polo_1\,^\mu \polo_3\,^\nu \right) \left( \polo_2 . \polo_4 \right)+\left( {\tilde \varepsilon}^2_{\mu\nu} 
\polo_2\,^\mu \polo_4\,^\nu \right) \left( \polo_1 . \polo_3 \right).
\eea
}

\subsubsection*{\underline{$D=4$}}
The parity odd module generators $O_{\bf 3}^{D=4}$ and $O_{\bf S}^{D=4}$ both need explicit $\Z_2\times \Z_2$ symmetrization. Projecting instead on the state with $(+--)$ charge,
\be\label{podfgen} 
\begin{split}
	{\tilde O}_1^{D=4,(1)}&\equiv 2*(F_1\wedge F_2){\rm Tr}(F_3 F_4)|_{(+--)}=4*(F_1\wedge F_2){\rm Tr}(F_3 F_4)-4*(F_3\wedge F_4){\rm Tr}(F_1 F_2),\nonumber\\
	{\tilde O}_2^{D=4,(1)}&\equiv 6\varepsilon_{\mu\nu\rho\sigma}F_1^{\mu\nu} \partial^\rho F_2^{ab}\partial^\sigma F_3^{bc} F_4^{ca}|_{(+--)}=6\big(\varepsilon_{\mu\nu\rho\sigma}F_1^{\mu\nu} \partial^\rho F_2^{ab}\partial^\sigma F_3^{bc} F_4^{ca}+ \varepsilon_{\mu\nu\rho\sigma}F_2^{\mu\nu} \partial^\rho F_1^{ab}\partial^\sigma F_4^{bc} F_3^{ca}\nonumber\\
	&- \varepsilon_{\mu\nu\rho\sigma}F_3^{\mu\nu} \partial^\rho F_4^{ab}\partial^\sigma F_1^{bc} F_2^{ca}-\varepsilon_{\mu\nu\rho\sigma}F_4^{\mu\nu} \partial^\rho F_3^{ab}\partial^\sigma F_2^{bc} F_1^{ca}\big).
\end{split}
\ee
Here the first structure comes from $O_{\bf 3}^{D=4}$ and the second comes from $O_{\bf S}^{D=4}$. Both are symmetric under $P_{12}$, hence transform as ${\bf 3}$. We denote these structures as ${\tilde O}_{{\bf 3},1}^{D=4}$ and ${\tilde O}_{{\bf 3},2}^{D=4}$. In summary, while the parity even quasi non-invariant generator remains unchanged (see \eqref{photon-lag2}),  in $D=4$ there are are two parity odd quasi invariant photon modules at fourth order in derivatives and they transform in ${\bf 3}$.

In summary, we tabulate the $S_3$ representation and the derivative order of non-quasi-invariant structures in table \ref{non-inv-photon-table}.
\begin{table}
	\begin{center}
		\begin{tabular}{|l|l|l|}
			\hline
			Parity even & $n_{\bf 3}$ & $n_{\bf 3_A}$  \\
			\hline
			$D\geq 4$ & $1_{6\partial}\qquad\,\,$ & $0$ \\
			\hline
		\end{tabular}\\
		\vspace{0.3cm}
		\begin{tabular}{|l|l|l|}
			\hline
			Parity odd & $n_{\bf 3}$ & $n_{\bf 3_A}$ \\
			\hline
			$D\geq 7$ & 0& 0 \\
			\hline
			$D=6$ & $0$ & $1_{4\partial}$  \\
			\hline
			$D=5$ & $1_{5\partial}$ & $2_{5\partial}$  \\
			\hline
			$D=4$ & $1_{4\partial}+ 1_{6\partial}$ & 0 \\
			\hline
		\end{tabular}
	\end{center}
	\caption{The counting of non-quasi-invariant photon structures along with their $S_3$ representation and derivative order. The counting for $D=5$ is the same as that for $D\geq 7$.}
	\label{non-inv-photon-table}
\end{table}

\subsubsection{Tensor product and projection}\label{non-inv-tensor}
We construct the submodule $\CM^{\rm non-inv}$ of gluon module by taking the tensor product of non-quasi-invariant photon structures (detailed in section \ref{non-inv-photon}) and non-quasi-invariant color structures (detailed in section \ref{non-inv-color}) and projecting onto $\Z_2\times \Z_2$ invariant states. We will not do this explicitly for all dimensions and for all gauge groups but rather illustrate the procedure for the case of $D\geq 7$ and gauge group $SU(N)$ for $N\geq 3$. In this case, we have the photon structures ${\tilde E}_{\bf 3}$ (see \eqref{photon-lag2}) and color structure ${\tilde \xi}_{\bf 3_A}$ (see \eqref{mnisun}). Projecting the tensor product onto $\Z_2\times \Z_2$ invariants, we get three states,
\be
({\tilde E}_{\bf 3}^{(1)} {\tilde \xi}_{\bf 3_A}^{(1)}, {\tilde E}_{\bf 3}^{(2)} {\tilde \xi}_{\bf 3_A}^{(2)}, {\tilde E}_{\bf 3}^{(3)} {\tilde \xi}_{\bf 3_A}^{(3)}).
\ee 
As the $P_{12}$ charge of ${\tilde E}_{\bf 3}^{(1)}$ is $+1$ and that of $ {\tilde \xi}_{\bf 3_A}^{(1)}$ is $-1$, the $P_{12}$ charge of their product is $-1$. This makes the above quasi-invariant generators transform in representation ${\bf 3_A}$ of $S_3$. More explicitly this module is generated by the following Lagrangian,
\bea \label{lagmnisun}
H^{1}_{SU(N)} &=& \sum_{m,n}  a_{m,n} \prod_{b=1}^m\prod_{c=1}^n {\rm Tr}(\{T^e, T^f \}[T^g ,T^h ] )~\left(\partial_{\m_b} \partial_{\n_c} F^e_{\a\b} \partial_{\a}F^f_{jk} \partial^{\m_b}\partial_{\b}F^g_{kl} \partial^{\n_c} F^h_{lj} \right. \nonumber\\
&&~~~~~~~~~\left.- \partial_{\m_b} \partial_{\n_c} F^f_{\a\b} \partial_{\a}F^e_{jk} \partial^{\m_b}\partial_{\b}F^g_{kl} \partial^{\n_c} F^h_{lj}\right).
\eea
We present the explicit generators of $\CM^{\rm non-inv}$ for $SO(N<9)$ and $SU(N<4)$ $D\geq8$ in appendix \ref{egmsnso} and \ref{egmsnsu}.

Finally, the partition function results about the generators of the $\CM^{\rm non-inv}$ for all the cases, including $D<8$, are tabulated in table \ref{M-non-inv-table}. We have encoded the $S_3$ representation and the derivative order of the gluon generators of $\CM^{\rm non-inv}$ by specifying the pair $(n_{\bf 3},n_{\bf 3_A})=(0,1_{6\partial})$. The contribution of the $\CM^{\rm non-inv}$ precisely matches with the underlined part of tables \ref{gluon-plethystic}, \ref{gluonplethd7}, \ref{gluonplethd6}, \ref{gluonplethd5} and \ref{gluonplethd4}. This confirms that our counting of gluon Lagrangians is complete.
\begin{table}
	\begin{center}
		\begin{tabular}{|l|l|l|l|}
			\hline
			SO(N) & $N\geq 7$ & $N=6$ & $N=4$ \\
			\hline
			$D\geq 7$ & (0,0) & (0,$1_{6\partial}$) & ($1_{6\partial}$,0)  \\
			\hline
			$D=6$ & (0,0) & ($1_{4\partial}$,$1_{6\partial}$) & ($1_{6\partial}$,$1_{4\partial}$)  \\
				\hline
			$D=5$ & (0,0) & ($2_{5\partial}$,$1_{6\partial}+1_{5\partial}$) & ($1_{6\partial}+1_{5\partial}$,$2_{5\partial}$)   \\
			\hline
			$D=4$ & (0,0) & (0,$1_{4\partial}+2_{6\partial}$) & ($1_{4\partial}+2_{6\partial}$,0)  \\
			\hline
		\end{tabular}
		\quad
		\begin{tabular}{|l|l|l|}
			\hline
			SU(N) & $N\geq 3$ & $N=2$  \\
			\hline
			$D\geq 7$ & ($1_{6\partial}$,0) & (0,0)  \\
			\hline
			$D=6$ & ($1_{4\partial}$,$1_{6\partial}$) & (0,0)  \\
			\hline
			$D=5$ & ($2_{5\partial}$,$1_{6\partial}+1_{5\partial}$) & (0,0)  \\
			\hline
			$D=4$ & (0,$1_{4\partial}+2_{6\partial}$) & (0,0) \\
			\hline
		\end{tabular}
	\end{center}
	\caption{The number of generators $(n_{\bf 3},n_{\bf 3_A})$ of $\CM^{\rm non-inv}$ along with their $S_3$ representation and the derivative order. }
	\label{M-non-inv-table}
\end{table}

\subsection{Yang Mills gluon amplitude}

As a check of our basis for gluon structures, we express the four gluon amplitude in terms of our structures. The four gluon amplitude from pure yang mills is given by \cite{Bern:2019prr}, 

\bea
i {\cal A}_4^{\rm{tree}}=g^2 \left( \frac{n_s c_s}{s} + \frac{n_t c_t}{t} + \frac{n_u c_u}{u} \right)
\eea                          

where
\bea
n_s &=&-\frac{1}{2}\{\left[ (\ve_1.\ve_2)p_1^{\mu} +2(\ve_1.p_2) \ve_2^\mu -(1 \leftrightarrow 2)\right]\left[ (\ve_3.\ve_4) p_{3\mu} +2(\ve_3.p_4) \ve_{4\mu} -(3 \leftrightarrow 4)\right] \nonumber\\
&&+s\left[(\ve_1.\ve_3)(\ve_2.\ve_4)- (\ve_1.\ve_4)(\ve_2.\ve_3)\right]\} \nonumber\\
c_s &=& -2f^{a_1a_2b} f^{ba_3a_4},\qquad c_un_u=c_sn_s|_{1\rightarrow 2\rightarrow 3\rightarrow 1}, \qquad c_tn_t=c_sn_s|_{1\rightarrow 3\rightarrow 2\rightarrow 1}
\eea                          
The $stu$ descendant of this S-matrix arises from the following Lagrangian with $G^i$ defined in equation \eqref{plagsonadj}.
\begin{eqnarray}
stu {\cal A}_4^{\rm{tree}} \propto \left(\left(G^4_{SU(N)}-\frac{G^5_{SU(N)}}{2}-\frac{G^6_{SU(N)}}{2}\right)-4\left( G^{10}_{SU(N)}-\frac{G^{11}_{SU(N)}}{2}-\frac{G^{12}_{SU(N)}}{2} \right) \right)|_{m=1,n=0}.\quad 
\end{eqnarray}                 

\section{Summary and outlook}\label{summary}
We have classified four-point local gluon S-matrices in arbitrary number of dimensions and  for gauge group $SO(N)$ and $SU(N)$. Our method is general and can be applied in the straightforward way to other gauge groups as well. As explained in \cite{Chowdhury:2019kaq}, the four-point local S-matrices are in one to one correspondence with Lagrangian terms that are quartic in corresponding fields modulo total derivatives and equation of motion. So, in effect, the classification of four-point S-matrices is equivalent to classification of equivalence classes of quartic Lagrangians i.e. Lagrangians containing four Field strength operators.  Our classification follows the same strategy as \cite{Chowdhury:2019kaq} where four-point photon and graviton S-matrices were classified. In particular we identify the generators (and also relations, in case there are any) of the module $\CM^{\rm gluon}$ of quasi-invariant\footnote{i.e. invariant under only $\Z_2\times \Z_2$ subgroup of the permutation group $S_4$.} local S-matrices. 

We have used the fact that the gluon S-matrix admits a type of ``factorization" into the S-matrix of adjoint scalars and that of photons. More precisely, 
\be\label{M-glue}
\CM^{\rm gluon} = \CM^{\rm inv} \oplus \CM^{\rm non-inv},\qquad \qquad {\rm where}\quad\CM^{\rm inv}= \CM^{\rm scalar} \otimes \CM^{\rm photon}.
\ee
The submodule $\CM^{\rm non-inv}$ is a projection of the tensor product of the scalar and photon part. Our classification is done by identifying all the individual scalar and photon components involved in equation \eqref{M-glue}. This construction is spelled out in detail in section \ref{explicit}. For the case of $D\geq 8$ and for $SO(N>9), SU(N>3)$, the explicit Lagrangians generating the gluon S-matrices are given in equation \eqref{plagsonadj}.
Moreover, we define a partition function over the space of local S-matrices
\be
Z(x)={\rm Tr} \,x^\partial
\ee
where $\partial$ is the derivative order. We have tabulated the partition functions in all the dimensions and for all $SO(N),SU(N)$ gauge groups in tables \ref{gluon-plethystic}, \ref{gluonplethd7}, \ref{gluonplethd6}, \ref{gluonplethd5} and \ref{gluonplethd4}. 

In past few years, starting with the works \cite{Bern:2008qj, Bern:2010ue, Bern:2010yg}, the gluon scattering has been used to compute graviton scattering amplitude. The relation between the two amplitudes goes by the name color/kinematic duality or the double copy relation or the BCJ relation. This relation is a generalization of the relation between the gluon scattering and graviton scattering in string theory discovered in mid-80s \cite{Kawai:1985xq}, the so-called KLT relation. In spirit, these relations follow due to a kind of factorization of the gluon S-matrix into the S-matrix of adjoint scalars and that of photons. The scalar S-matrix keeps track of the color structure while the photon S-matrix keeps track of the polarizations and momenta. The BCJ relation proposes replacing the scalar part i.e. color structure in the gluon S-matrix in Yang-Mills theory  by another copy of the photon part i.e. the kinematic structure to obtain graviton S-matrix in Einstein gravity. In our classification also, the factorized structure i.e. equation \eqref{M-glue} plays an important role. It is tempting to guess that it may lead to a generalization of the BCJ relation between gauge theories different from Yang-Mills and gravitational theories different from Einstein gravity (See  \cite{Johansson:2017srf, Johansson:2018ues} for generalising colour/kinematics duality to supergravity amplitudes, \cite{Huang:2013kca, Huang:2012wr} for colour/kinematics duality in the context of  ABJM theories and \cite{,Broedel:2012rc, Engelund:2012re} for colour/kinematics duality in the context of effective field theories).

The index structures of the gluon S-matrices that we have classified also serves as the classification for index structures of four-point function of non-abelian currents in conformal field theory. These structures have   been studied in \cite{Costa:2011mg, Costa:2011dw, Kravchuk:2016qvl}. In the same way, the local scalar S-matrices give rise to the so called ``truncated solutions"\footnote{i.e. solutions having support over only finitely many spins} of the conformal crossing equation \cite{Heemskerk:2009pn}, we expect the gluon S-matrices to parametrize the truncated solutions for the crossing equation of non-abelian currents. It would be nice to explore this possibility further.
Finally, it would be interesting to classify higher point scattering of gluons and study the interplay of the index structures with the BCFW recursion relation \cite{Britto:2005fq}.

\acknowledgments
We thank Mrunmay Jagadale and Trakshu Sharma for the collaboration in the initial stages of the project. 
We would like to thank Alok Laddha, R. Loganayagam, Shiraz Minwalla, Siddharth Prabhu and Sandip Trivedi  for interesting questions and discussions on the subject. 
The work of both authors was supported by the Infosys Endowment for the study of the Quantum Structure of Spacetime.
The work of A.G. is also supported by SERB Ramanujan fellowship. We would all also like to acknowledge our debt to the people of India for their steady support to the study of the basic sciences.

\appendix
\section{Symmetric group $S_3$ and its representations}\label{s3}

The symmetric group $S_3$ is the group of symmetries of an equilateral triangle. It consists of rotating the triangle by $2\pi/3$ which generates the subgroup $\Z_3$ and also of the reflection across the central axis, this generates the subgroup $\Z_2$. The two subgroups $\Z_3$ and $\Z_2$ don't commute. However, $\Z_3$ is a normal subgroup which makes $S_3$ a semi-direct product $\Z_3\ltimes \Z_2$.

In order to construct irreducible representations of $S_3$, we first diagonalize $\Z_2$. The irrep is then labeled by the $\Z_2$ eigenvalue $+1$ or $-1$ and it is an irrep of $\Z_3$. Consider the $N$ dimensional representation of $\Z_N$ where $\Z_N$ acts by cyclic permutation. This is a reducible representation. The sum of all the elements is an invariant under $\Z_N$ action, hence $N$ dimensional representation decomposes as $1\oplus (N-1)$. The $N-1$ dimensional representation is the representation of $N$ elements that get cyclically permuted but sum to $0$.
In the case of $S_3$, we can consider the natural $3$ dimensional of $\Z_3$. It is $\Z_2$ even, we denote it as ${\bf 3}$ and if it is $\Z_2$ odd we denote it as ${\bf 3_A}$. As remarked earlier, both these representations are reducible. 
\be
{\bf 3}={\bf 1_S}\oplus {\bf 2_M},\qquad {\bf 3_A}={\bf 1_A}\oplus {\bf 2_M}.
\ee
Here ${\bf 1}_S$ is a one dimensional representation that is invariant under $\Z_3$ as well as $\Z_2$ and hence invariant under the whole $S_3$. The subscript ${\bf S}$ stands for symmetric.  The representation ${\bf 1}_A$ is a one dimensional representation that is invariant under $\Z_3$ but odd under $\Z_2$. The subscript ${\bf A}$ stands for anti-symmetric. The only other irreducible representation is ${\bf 2_M}$. Here the subscript $\bf M$ stands for mixed. In terms of standard Young diagrams, these representations are
\be
{\bf 1_S}=\yng(3),\qquad {\bf 2_M}=\yng(2,1),\qquad {\bf 1_A}=\yng(1,1,1).
\ee

 \subsection*{Clebsch-Gordon rules}\label{cgrules}
 Clearly
 \begin{equation}\label{cgoeff} 
 {\bf 1_S} \otimes {\bf R}={\bf R} 
 \end{equation}
 where ${\bf R}$ denotes any of the irreps   ${\bf 1_S}$, ${\bf 1_A}$ or ${\bf 2_M}$.
 On the other hand we have
 \begin{equation}\label{cgoeff1} 
 {\bf 1_A} \otimes {\bf 1_S}={\bf 1_A},~~~
 {\bf 1_A} \otimes {\bf 1_A}={\bf 1_S}, ~~~
 {\bf 1_A} \otimes {\bf 2_M}={\bf 2_M}. 
 \end{equation}
 Finally we have 
 \begin{equation} \label{ttt}
 {\bf 2_M } \otimes {\bf 2_M}= {\bf 1_S} \oplus {\bf 1_A} \oplus {\bf 2_M}.
 \end{equation}
 
\section{Projectors on the tensor product}\label{projectors}
In this section, we derive the projectors \eqref{proj3} and \eqref{proj4} of subsection \ref{count-inv}. Let us denote the states in the tensor product of four identical $G$-representations $\rho$ by $| \a_1, \a_2, \a_3, \a_4 \rangle$. The character of the representation due to the anti-symmetric projector \eqref{proj3} is given by
\bea\label{proj3calc}
&&\chi_{\wedge^4\rho\,}(a_i)\nonumber\\
&=&\sum_{\alpha_1 \alpha_2 \a_3 \a_4} \langle \alpha_1,\alpha_2, \a_3, \a_4 | a_i^{H_i} \left(\frac{1-P_{12}-P_{23}-P_{13}+P_{23}P_{12}+P_{12}P_{23}}{6}\right)\left(\frac{(1+P_{12}P_{34}+P_{13}P_{24}+P_{14}P_{23})}{4}\right)\nonumber\\
&&|\alpha_1,\alpha_2, \a_3, \a_4 \rangle,\nonumber\\
&=&\frac{1}{24}\left( \chi_\rho(a_i)^4 +3\chi_\rho(a_i^2)^2-6\chi_\rho(a_i)^2\chi_\rho(a_i^2)-\chi_\rho(a_i^4)+8\chi_\rho(a_i^3)\chi_\rho(a_i)\right) \nonumber\\
&=& \chi_{\wedge^2\rho}(a_i)^2 - \chi_{\wedge^2\rho}(a_i) \otimes \chi_{S^2\rho}(a_i) - \chi_{S^2\rho}(a_i)^2 - \chi_{S^4\rho}(a_i) + 2 \chi_{S^3\rho}(a_i) \otimes \chi_{\rho}(a_i)
\eea
where we have used the following identities to eliminate $\chi_\rho(a_i^4), \chi_\rho(a_i^3), $ and $\chi_\rho(a_i^2)$,  
\bea\label{projidentity}
\chi_\rho(a_i^4)&=&4 \chi_{S^4 \rho}(a_i)- \frac{1}{6}\left( \chi_\rho(a_i)^4+3\chi_\rho(a_i^2)^2+6\chi_\rho(a_i)^2\chi_\rho(a_i^2)+8\chi_\rho(a_i^3)\chi_\rho(a_i) \right)\nonumber\\
\chi_\rho(a_i^3)&=&3 \chi_{S^3 \rho}(a_i)- \frac{1}{2}\left( \chi_\rho(a_i)^3+3\chi_\rho(a_i^2)\chi_\rho(a_i)\right)\nonumber\\
\chi_\rho(a_i^2)&=&\left( \chi_{S^2 \rho}(a_i)- \chi_{\wedge^2 \rho}(a_i)\right)\nonumber\\
\chi_\rho(a_i)^2&=&\left( \chi_{S^2 \rho}(a_i)+ \chi_{\wedge^2 \rho}(a_i)\right).
\eea 

The character of the representation due to the $\rho_{{\bf 3}}$ projector \eqref{proj4} is given by
\bea\label{proj4calc}
\chi_{\rho_{{\bf 3}}\,}(a_i)
&=&\sum_{\alpha_1 \alpha_2 \a_3 \a_4} \langle \alpha_1,\alpha_2, \a_3, \a_4 | a_i^{H_i} \,\left(\frac{1+ P_{12}}{2}\right)\left(\frac{(1+P_{12}P_{34}-P_{13}P_{24}-P_{14}P_{23})}{4}\right)|\alpha_1,\alpha_2, \a_3, \a_4 \rangle,\nonumber\\
&=&\frac{1}{8}\left( \chi_\rho(a_i)^4 -\chi_\rho(a_i^2)^2+ 2\chi_\rho(a_i)^2\chi_\rho(a_i^2)-2\chi_\rho(a_i^4)\right) \nonumber\\
&=&- \chi_{S^4\rho}(a_i) + \chi_{S^3\rho}(a_i) \otimes \chi_{\rho}(a_i)
\eea
where we have used the identities \eqref{projidentity}. Similarly character of the representation due to the $\rho_{{\bf 3_A}}$ projector is given by 
\bea\label{proj4calc2}
\chi_{\rho_{{\bf 3_A}}\,}(a_i)&=&\sum_{\alpha_1 \alpha_2 \a_3 \a_4} \langle \alpha_1,\alpha_2, \a_3, \a_4 | a_i^{H_i} \,\left(\frac{1- P_{12}}{2}\right)\left(\frac{(1+P_{12}P_{34}-P_{13}P_{24}-P_{14}P_{23})}{4}\right)|\alpha_1,\alpha_2, \a_3, \a_4 \rangle,\nonumber\\
&=&\frac{1}{8}\left( \chi_\rho(a_i)^4 -\chi_\rho(a_i^2)^2-2\chi_\rho(a_i)^2\chi_\rho(a_i^2)+2\chi_\rho(a_i^4)\right) \nonumber\\
&=& \chi_{S^4\rho}(a_i) - \chi_{S^3\rho}(a_i) \otimes \chi_{\rho}(a_i)+\chi_{S^2\rho}(a_i) \otimes \chi_{\wedge^2\rho}(a_i)
\eea
Equation \eqref{z2z2invk} can be verified in a similar manner. 

\section{ Plethystic integrals}\label{pi}
In this section we provide the details to perform the plethystic integrals in section \ref{plethystic}. 
The haar measure for $SO(D)$ is given by 
\be
d\mu_{SO(D)}=\prod_{i=1}^{\lfloor D/2\rfloor}dy_i \,\Delta(y_i)\,
\ee 
where, for even dimensions ($D=2N$), the Vandermonde determinant for $SO(D)$ is given by 
\begin{equation}\label{haare}
\Delta_e(y_i) =\frac{2 \left(\prod _{j=1}^{N} \left(\prod _{i=1}^{j-1} \left(y_i+\frac{1}{y_i}-y_j-\frac{1}{y_j}\right)\right)\right)^2}{(2\pi i)^N 2^NN!\prod_{i=1}^{N}y_i},
\end{equation} 
and for odd dimensions ($D=2N+1$), the Vandermonde determinant is given by                         
\begin{equation}\label{haaro}
\Delta_o(y_i) =\frac{\left(\prod _{k=1}^N \left(1-y_k-\frac{1}{y_k}\right)\right) \left(\prod _{j=1}^N \left(\prod _{i=1}^{j-1}  \left(y_i+\frac{1}{y_i}-y_j-\frac{1}{y_j}\right)\right)\right)^2}{(2\pi i)^N N!\prod_{i=1}^{N}y_i}.
\end{equation}
The integral over $y_i$ in \eqref{singlet-proj} is a closed circular contour about $y_i=0$.

The Haar measure for $SU(N)$ is given by \cite{Gray:2008yu}, 
\be\label{haarsu}
d\mu_{SU(N)}=\frac{1}{(2\pi i)^{(N-1)} N!} \prod_{l=1}^{N-1} \frac{dz_l}{z_l} \D(\phi)\D(\phi^{-1})	
\ee 
where $\phi_a(z_1,\ldots z_{N-1})|_{a=1}^N$ are the coordinates on the maximal torus of $SU(N)$ with $\prod_{l=1}^N \phi_l=1$ and $\D(\phi)=\prod_{1\leq a< b\leq N} (\phi_a-\phi_b)$ is the Vandermonde determinant and the integral over $z_i$ in \eqref{singlet-proj} is a closed circular contour about $z_i=0$. Explicitly written out, the coordinates on the maximal torus take the form, 
\be\label{coordmaxtor}
\phi_1=z_1,\qquad \phi_k=z^{-1}_{k-1}z_k,\qquad \phi_N=z^{-1}_{N-1}.
\ee

The integrals at hand, \eqref{singlet-proj} and \eqref{singlet-projgluon}, therefore have two Haar integrals one of which pertains to projecting onto Lorentz singlets, while the other is to project onto the colour singlets. We perform the Haar integral for the Lorentz singlets first. We note that the Haar integral for the Lorentz group stabilizes for $D>3$ for scalars and $D>8$ for photons. 
Using large $D$ techniques of \cite{Chowdhury:2019kaq} we obtain, for scalars 
\begin{eqnarray}
\oint \prod_{i=1}^{\lfloor D/2\rfloor}dy_i \,\Delta(y_i)\,  i_s^{4}(x,y)/\denom(x,y) &=& \frac{1}{\left(x^2-1\right)^2}\nonumber\\
\oint \prod_{i=1}^{\lfloor D/2\rfloor}dy_i \,\Delta(y_i)\,  i_s^{2}(x,y)i_s(x^2,y^2)/\denom(x,y) &=& \frac{1}{1-x^4}\nonumber\\
\oint \prod_{i=1}^{\lfloor D/2\rfloor}dy_i \,\Delta(y_i)\,  i_s(x^2,y^2)^2/\denom(x,y) &=& \frac{1}{\left(x^2-1\right)^2}\nonumber\\
\oint \prod_{i=1}^{\lfloor D/2\rfloor}dy_i \,\Delta(y_i)\,  i_s(x^3,y^3)i_s(x,y)/\denom(x,y) &=& \frac{1}{\left(x^4+x^2+1\right)}\nonumber\\
\oint \prod_{i=1}^{\lfloor D/2\rfloor}dy_i \,\Delta(y_i)\,  i_s(x^4,y^4)i_s(x,y)/\denom(x,y) &=& \frac{1}{\left(1-x^4\right)}
\end{eqnarray}  
and for photons
\begin{eqnarray}
\oint \prod_{i=1}^{\lfloor D/2\rfloor}dy_i \,\Delta(y_i)\,  i_v^{4}(x,y)/\denom(x,y) &=& \frac{2x^8 \left(2 x^2+3\right)}{\left(x^2-1\right)^2}\nonumber\\
\oint \prod_{i=1}^{\lfloor D/2\rfloor}dy_i \,\Delta(y_i)\,  i_v^{2}(x,y)i_v(x^2,y^2)/\denom(x,y) &=& \frac{2x^8}{1-x^2}\nonumber\\
\oint \prod_{i=1}^{\lfloor D/2\rfloor}dy_i \,\Delta(y_i)\,  i_s(x^2,y^2)^2/\denom(x,y) &=& \frac{6 x^8}{ \left(x^2-1\right)^2}\nonumber\\
\oint \prod_{i=1}^{\lfloor D/2\rfloor}dy_i \,\Delta(y_i)\,  i_v(x^3,y^3)i_v(x,y)/\denom(x,y) &=& \frac{x^{10}}{ \left(x^4+x^2+1\right)}\nonumber\\
\oint \prod_{i=1}^{\lfloor D/2\rfloor}dy_i \,\Delta(y_i)\,  i_v(x^4,y^4) &=& \frac{2x^8}{1-x^4}.
\end{eqnarray}  
The integrals \eqref{singlet-proj} and \eqref{singlet-projgluon} then take the schematic form, 
\be\label{scalar-proj-colour2}
\begin{split}
	I^D_\ts(x):= \oint d\mu_{G}~&\left(\frac{\chi^{G}_R(z^2) \chi^{G}_R(z)^2}{4 \left(1-x^4\right)}+\frac{\chi^{G}_R(z^4)}{4 \left(1-x^4\right)}+\frac{\chi^{G}_R(z)^4}{24 \left(x^2-1\right)^2}+\frac{\chi^{G}_R(z^2)^2}{8 \left(x^2-1\right)^2}\right.\\
	&\left. +\frac{\chi^{G}_R(z^3)\chi^{G}_R(z)}{3 \left(x^4+x^2+1\right)}\right)\\
	:=\oint d\mu_{G}~& I^N_s(G)\\
I_\tv^{D}(x,y,z):=\oint  d\mu_{G}~& \left(\frac{x^8 \chi^{R}_a(z^4)}{2-2 x^4} +\frac{\left(x^8 \left(2 x^2+3\right)\right) \chi^{R}_a(z)^4}{12 \left(x^2-1\right)^2}+\frac{\left(3 x^8\right) \chi^{R}_a(z^2)^2}{4 \left(x^2-1\right)^2} \right.\\
&\left.~~~~~~~~~~ +\frac{x^8 \chi^{R}_a(z)^2 \chi^{R}_a(z^2)}{2-2 x^2}+\frac{x^{10} \chi^{R}_a(z) \chi^{R}_a(z^3)}{3 \left(x^4+x^2+1\right)}\right)\\
:=\oint d\mu_{G}~& I^N_\tv(G).
\end{split}
\ee

\subsection{$SO(N)$}\label{pison}
When the colour group $G$ is $SO(N)$, we do the haar integral in the following manner. For $N\geq 9$, we do a large $N$ integral following \cite{Chowdhury:2019kaq}.  
\be \label{soadj}
\begin{split}
	\oint \prod_{i=1}^{\lfloor N/2\rfloor}dy_i \,\Delta(y_i)\,~ \chi^{SO(N)}_a( y^2)\chi^{SO(N)}_a( y)^2 &=2\\
	\oint \prod_{i=1}^{\lfloor N/2\rfloor}dy_i \,\Delta(y_i)\, ~\chi^{SO(N)}_a( y^4) &=2\\
	\oint \prod_{i=1}^{\lfloor N/2\rfloor}dy_i \,\Delta(y_i)\, ~\chi^{SO(N)}_a( y)^4 &=6\\
	\oint \prod_{i=1}^{\lfloor N/2\rfloor}dy_i \,\Delta(y_i)\, ~\chi^{SO(N)}_a( y^2)^2 &=6\\
	\oint \prod_{i=1}^{\lfloor N/2\rfloor}dy_i \,\Delta(y_i)\, ~\chi^{SO(N)}_a( y^3)\chi^{SO(N)}_a( y) &=0
\end{split}
\ee 
where  
\bea 
\chi^{SO(N)}_a( y) &=& \left( \frac{\chi^{SO(N)}_f(y)^2-\chi^{SO(N)}_f(y^2)}{2} \right)\nonumber\\
\chi^{SO(N)}_f(y) &=& \sum_{i=1}^{N/2} \left( y_i + \frac{1}{y_i}\right)\qquad \qquad \qquad \qquad{\rm for \,\, N\,\, even}\nonumber \\
&=& \sum_{i=1}^{\lfloor N/2 \rfloor} \left( y_i + \frac{1}{y_i}\right)+1\qquad \qquad \qquad {\rm for \,\, N\,\, odd}.
\eea
The final result for projection onto colour singlets become,
\be \label{partfnsonadj2}
\begin{split} 
I^{a}_{\ts,~ SO(N)}(x) &= \frac{2+2x^2+2x^4}{(1-x^4)(1-x^6)} =2Z_{{\bf 3}}\\
I^{a}_{\tv,~ SO(N)}(x) &= \frac{2x^4(4+7x^2+7x^4+3x^6)}{(1-x^4)(1-x^6)} =x^4(8Z_{{\bf 3}}+4Z_{{\bf 3_A}}+2x^2 Z_{{\bf 3}}).
\end {split}
\ee 
For $N\leq 8$, we resort to numerical integration. We do not expound the details. Interested readers may consult Appendix H.2 of \cite{Chowdhury:2019kaq} for details on how to perform the numerical integration. The results for the numerical integration are recorded in  table \ref{scalar-plethystic} for scalars and table \ref{gluon-plethystic} for gluons. Note that the relevant Lorentz haar integrals for $D< 8$ (in tables \ref{gluonplethd7}, \ref{gluonplethd6}, \ref{gluonplethd5}, \ref{gluonplethd4}) have been done numerically in a similar manner.

\subsection{$SU(N)$}\label{pisun}
For $SU(N)$ the Haar integral stabilises for $N \geq 4$. We verify this from an explicit large $N$ computation of the plethystic integral \eqref{scalar-proj-colour2}. We consider the case when fields are charged under the adjoint representation of $SU(N)$. For $N<4$, we resort to numerical integration. 

\subsubsection{$N\geq 4$}\label{pisunlargen}   
In this subsubsection we perform the large $N$ integral for $SU(N)$. The Haar measure exponentiates as follows
\begin{eqnarray}
&&\Pi_{1\leq a<b\leq N} \left(\phi_a-\phi_b\right) \Pi_{1\leq m<p\leq N}\left(\phi_m^{-1}-\phi_p^{-1}\right)\nonumber\\
&&\propto \Pi_{1\leq a<b\leq N} \left(1-\frac{\phi_b}{\phi_a}\right) \Pi_{1\leq m<p\leq N}\left(1-\frac{\phi_m}{\phi_p}\right)\nonumber\\
&&\propto e^{-\sum_n \frac{1}{n}\left(\sum_{1\leq a<b\leq N} \left( \frac{\phi_b}{\phi_a}\right)^n +\sum_{1\leq m<p\leq N} \left( \frac{\phi_m}{\phi_p}\right)^n \right)} \nonumber\\
&&\propto e^{-\sum_n \frac{1}{n}\left(\chi^{SU(N)}_{f}\left( \phi^n\right)\chi^{SU(N)}_{\bar{f}}\left( \phi^n\right) -1\right)} \nonumber\\
&&\propto e^{-\sum_n \frac{1}{n}\left({\rm Tr}U^n{\rm Tr}U^{-n} -1\right)} \nonumber\\
&&\propto e^{-\sum_n \frac{1}{n}\chi^{SU(N)}_a \left(\phi^n\right)}
\end{eqnarray}                     
where $U$ is an $N \times N$ diagonal matrix
$$ U \sim\left( \begin{matrix}
\phi_1 & 0 &0 &\cdots &0  \\
0 & \phi_2 &0 &\cdots &0\\
0 & 0 & \phi_3 &\cdots &0 \\
\cdots&\cdots& \cdots& \cdots&\cdots\\
0 & 0 &\cdots &\cdots &\phi_N\\
\end{matrix}\right)$$
and we have used the definition of the character of the fundamental and anti-fundamental representation     
\bea
\chi^{SU(N)}_f\left(\phi\right)&=& \sum_{i=1}^N \phi_i, \qquad
\chi^{SU(N)}_{\bar{f}}\left( \theta\right)= \sum_{i=1}^N \phi_i^{-1}  \nonumber\\
\chi^{SU(N)}_a \left(\phi \right) &=& \chi^{SU(N)}_f\left(\phi \right) \chi^{SU(N)}_{\bar{f}} \left(\phi \right) -1
\eea   
In the large $N$ limit, both ${\rm Tr}U^n$ and ${\rm Tr}U^{-n}$ scale as $O(N)$. We, therefore, work with order unity variables $\rho_n$ given by
\bea
{\rm Tr}U^n=N \rho_n, \qquad {\rm Tr}U^{-n}=N \rho_n^\dagger
\eea
The haar integral \eqref{scalar-proj-colour2} becomes,
\begin{eqnarray}\label{largensunhaar}
I_\ts^D(x) &=& \frac{\int {\cal D}\rho_n {\cal D}\rho^\dagger_n~ e^{-\frac{1}{n}N^2|\rho_n|^2}I^N_s(\rho_n)}{\int {\cal D}\rho_n {\cal D}\rho^\dagger_n~ e^{-\frac{1}{n}N^2|\rho_n|^2}}
\end{eqnarray}
where the integrand $I^N_s(\rho_n)$ in the large $N$ limit scales as ,
\begin{eqnarray}\label{insr}
I^N_s(\rho_n)&=&\left(\frac{\chi^{SU(N)}_a(\phi^2) \chi^{SU(N)}_a(\phi)^2}{4 \left(1-x^4\right)}+\frac{\chi^{SU(N)}_a(\phi^4)}{4 \left(1-x^4\right)}+\frac{\chi^{SU(N)}_a(\phi)^4}{24 \left(x^2-1\right)^2}+\frac{\chi^{SU(N)}_a(\phi^2)^2}{8 \left(x^2-1\right)^2}+\frac{\chi^{SU(N)}_a(\phi^3)\chi^{SU(N)}_a(\phi)}{3 \left(x^4+x^2+1\right)}\right)\nonumber\\
&=&\left(\frac{(N^2|\rho_2|^2-1)(N^2|\rho_1|^2-1)^2}{4 \left(1-x^4\right)}+\frac{(N^2|\rho_4|^2-1)}{4 \left(1-x^4\right)}+\frac{(N^2|\rho_1|^2-1)^4}{24 \left(x^2-1\right)^2}+\frac{(N^2|\rho_2|^2-1)^2}{8 \left(x^2-1\right)^2}\right.\nonumber\\
&&\left. +\frac{(N^2|\rho_1|^2-1)(N^2|\rho_3|^2-1)}{3 \left(x^4+x^2+1\right)}\right).
\end{eqnarray}
Each of the $\rho_n^\alpha$ integrals in \eqref{largensunhaar} can be written in terms of a gaussian integral. For example take the second term in \eqref{insr}
\begin{eqnarray}
\frac{\int {\cal D}\rho_n {\cal D}\rho^\dagger_n~ e^{-\frac{1}{n}N^2|\rho_n|^2}(N^2|\rho_4|^2-1)}{\int {\cal D}\rho_n {\cal D}\rho^\dagger_n~ e^{-\frac{1}{n}N^2|\rho_n|^2}}&=& \frac{\int {\cal D}\rho_4 {\cal D}\rho^\dagger_4~ e^{-\frac{1}{4}N^2|\rho_4|^2}(N^2|\rho_4|^2-1)}{\int {\cal D}\rho_4 {\cal D}\rho^\dagger_4~ e^{-\frac{1}{4}N^2|\rho_4|^2}} \nonumber\\
&=& 3.
\end{eqnarray}     
All the terms in \eqref{insr} can be evaluated in this manner.
\begin{eqnarray}\label{largenintsun}
\frac{\int {\cal D}\rho_n {\cal D}\rho^\dagger_n~ e^{-\frac{1}{n}N^2|\rho_n|^2}(N^2|\rho_2|^2-1)(N^2|\rho_1|^2-1)^2}{\int {\cal D}\rho_n {\cal D}\rho^\dagger_n~ e^{-\frac{1}{n}N^2|\rho_n|^2}}&=& 1\nonumber\\
\frac{\int {\cal D}\rho_n {\cal D}\rho^\dagger_n~ e^{-\frac{1}{n}N^2|\rho_n|^2}(N^2|\rho_4|^2-1)}{\int {\cal D}\rho_n {\cal D}\rho^\dagger_n~ e^{-\frac{1}{n}N^2|\rho_n|^2}}&=& 3\nonumber\\
\frac{\int {\cal D}\rho_n {\cal D}\rho^\dagger_n~ e^{-\frac{1}{n}N^2|\rho_n|^2}(N^2|\rho_1|^2-1)^4}{\int {\cal D}\rho_n {\cal D}\rho^\dagger_n~ e^{-\frac{1}{n}N^2|\rho_n|^2}}&=& 9\nonumber\\
\frac{\int {\cal D}\rho_n {\cal D}\rho^\dagger_n~ e^{-\frac{1}{n}N^2|\rho_n|^2}(N^2|\rho_2|^2-1)^2}{\int {\cal D}\rho_n {\cal D}\rho^\dagger_n~ e^{-\frac{1}{n}N^2|\rho_n|^2}}&=& 5\nonumber\\
\frac{\int {\cal D}\rho_n {\cal D}\rho^\dagger_n~ e^{-\frac{1}{n}N^2|\rho_n|^2}(N^2|\rho_1|^2-1)(N^2|\rho_3|^2-1)}{\int {\cal D}\rho_n {\cal D}\rho^\dagger_n~ e^{-\frac{1}{n}N^2|\rho_n|^2}}&=& 0.
\end{eqnarray}
Putting \eqref{largensunhaar}, \eqref{insr}, \eqref{largenintsun} together we have 
\begin{eqnarray}
I_{\ts,~ SU(N)}^D(x) &=&  \frac{2(1+x^2+x^4)}{(1-x^4)(1-x^6)} =2Z_{{\bf 3}}\nonumber\\
I_{\tv,~ SU(N)}^D(x) &=& \frac{ x^4 (8 + 14 x^2 + 15 x^4 + 7 x^6+ x^{8})}{(1-x^4)(1-x^6)}=x^4(8Z_{{\bf 3}}+ 4Z_{{\bf 3_A}}+2x^2Z_{{\bf 3}}+x^2Z_{{\bf 3_A}}).
\end{eqnarray}     

\subsubsection{$N<4$}\label{pisunsmalln}
For $N<4$, we resort to numerical integration. As in \cite{Chowdhury:2019kaq}, we make the change of variables 
$$z_i = e^{i\theta_i}$$
The the contour integral over $z_i$ in \eqref{singlet-proj} becomes an angular integral over $\theta_i\sim (0 , 2\pi)$. The coordinates on the maximal torus  \eqref{coordmaxtor} become,
\be\label{coordmaxtorth}
\phi_1=e^{i \theta_1},\qquad \phi_k=e^{i (\theta_k- \theta_{k-1})},\qquad \phi_N=e^{-i \theta_{N-1}}
\ee
and the Haar measure \eqref{haarsu} becomes
\be\label{haarsuth}
d\mu_{SU(N)}=\frac{1}{(2\pi)^{(N-1)} N!} \prod_{l=1}^{N-1} d\theta_l \D(\phi)\D(\phi^{-1})	
\ee 
The integral for projecting onto colour singlets \eqref{scalar-proj-colour} is modified to 
\be\label{scalar-proj-coloursuadj}
\begin{split}
	I^a_{\ts, SU(N)}(x)&:= \int_0^{2\pi}\ldots\int_0^{2\pi} ~\frac{1}{(2\pi)^{(N-1)} N!} \prod_{l=1}^{N-1} d\theta_l \D(\phi)\D(\phi^{-1})\nonumber\\
	&~\left(\frac{\chi^{SU(N)}_a( e^{2i\theta_i}) \chi^{SU(N)}_a( e^{i\theta_i})^2}{4 \left(1-x^4\right)}+\frac{\chi^{SU(N)}_a( e^{4i\theta_i})}{4 \left(1-x^4\right)}+\frac{\chi^{SU(N)}_a( e^{i\theta_i})^4}{24 \left(x^2-1\right)^2}\right. \nonumber\\
	&\left. +\frac{\chi^{SU(N)}_a( e^{2i\theta_i})^2}{8 \left(x^2-1\right)^2}+\frac{\chi^{SU(N)}_a( e^{3i\theta_i})\chi^{SU(N)}_a( e^{i\theta_i})}{3 \left(x^4+x^2+1\right)}\right)\nonumber\\
\end{split}
\ee 
where
\bea
\chi^{SU(N)}_f\left(\theta\right)&=& e^{i \theta_1} + \sum_{i=2}^{N-1} e^{i (\theta_{i}-\theta_{i-1})}  + e^{-i \theta_{N-1}} \nonumber\\
\chi^{SU(N)}_{\bar{f}}\left( \theta\right)&=& e^{-i \theta_1} + \sum_{i=2}^{N-1} e^{-i (\theta_{i}-\theta_{i-1})}  + e^{i \theta_{N-1}} \nonumber\\
\chi^{SU(N)}_a \left(\theta \right) &=& \chi^{SU(N)}_f\left(\theta \right) \chi^{SU(N)}_{\bar{f}} \left(\theta \right) -1
\eea 
In order to perform this integral, we follow \cite{Chowdhury:2019kaq}. Assuming that the final answer can be reproduced as a sum over $Z_{{\bf 1_S}}$, $Z_{{\bf 2_M}}$ and $Z_{{\bf 1_A}}$, we multiply the 
plethystic integrand by $1/\denom$, Taylor series expand the 
result in $x$ around $0$. The coefficient of every power of $x$ in the result is an integral over $\theta_i$. 
We then evaluate these numerically using the Gauss-Kronrod method. As the numerical integration procedure is very accurate and can be performed very rapidly, we are able to perform this integral up to $x^{15}$ for scalars and $x^{30}$ for gluons. We thus able to verify that the polynomials in $x$ are finite (they terminate). The results for the numerical integration are recorded in table \ref{scalar-plethystic} for scalars and table \ref{gluon-plethystic} for gluons. Note that the relevant Lorentz haar integrals for $D< 8$ (in tables \ref{gluonplethd7}, \ref{gluonplethd6}, \ref{gluonplethd5}, \ref{gluonplethd4}) have been done numerically in a similar manner.

\section{Explicit Gluon module for small $N$ ($D \geq 9$)}\label{explicit-gluon-lag}
 In this appendix, we list the $D \geq 8$ ~$\CM^{\rm inv}$  and $\CM^{\rm non-inv}$ gluon modules for $SO(N<9)$ and $SU(N<4)$. 
\subsection{$SO(N)$}\label{egmsnso}
\subsubsection{$N=8$}
     There is an extra quasi invariant colour structure for $SO(8)$, listed in \eqref{qiso8}.	
$$\chi_{\bf S}^{SO(8)}=\Phi_1\wedge \Phi_2\wedge\Phi_3 \wedge \Phi_4.$$

This structure is $\Z_2 \otimes \Z_2$ symmetric and transform in ${\bf 1_S}$ of $S_3$. The tensor product of the quasi invariant colour and photon modules therefore generates $({\bf 3}\oplus{\bf 3})_{\rm photon}\otimes ({\bf 1_S})_{\rm scalar}=2 \cdot {\bf 3}$ generators at 4 derivatives and $({\bf 1_S})_{\rm photon}\otimes ({\bf 1_S})_{\rm scalar}= {\bf 1_S}$ generators at $6$ derivatives. They are generated by the folowing local Lagrangians. 
	\begin{eqnarray}\label{plagso8adj}
	G^{15}_{SO(8)} &=& \sum_{m,n}  a_{m,n} \prod_{b=1}^m\prod_{c=1}^n \ve^{\a\b\g\d\rho\s\xi\zeta}~T^e_{\a\b} T^f_{\g\d} T^g_{\rho\s} T^h_{\xi\zeta} 
	~\left(\partial_{\m_b} \partial_{\n_c} F^e_{ij} F^f_{ji} \right)  \left(\partial^{\m_b} F^g_{kl} \partial^{\n_c} F^h_{lk} \right) \nonumber\\ 
	G^{16}_{SO(8)} &=& \sum_{m,n}  a_{m,n} \prod_{b=1}^m\prod_{c=1}^n \ve^{\a\b\g\d\rho\s\xi\zeta}~T^e_{\a\b} T^f_{\g\d} T^g_{\rho\s} T^h_{\xi\zeta} 
	~\left(\partial_{\m_b} \partial_{\n_c} F^e_{ij} \partial^{\m_b}F^f_{jk} F^g_{kl} \partial^{\n_c} F^h_{li} \right) \nonumber\\
	G^{17}_{SO(8)} &=& \sum_{m,n}  a_{m,n} \prod_{b=1}^m\prod_{c=1}^n \ve^{\a\b\g\d\rho\s\xi\zeta}~T^e_{\a\b} T^f_{\g\d} T^g_{\rho\s} T^h_{\xi\zeta}  
	~\left(\partial_{\m_b} \partial_{\n_c} F^e_{ab} \partial_{a}F^f_{jk} \partial^{\m_b}\partial_{b}F^g_{kl} \partial^{\n_c} F^h_{lj} \right).
	\end{eqnarray}
where we have explained the notation in detail below \eqref{plagsonadj}. In summary for $SO(8)$, the local modules $\CM^{\rm inv}$ are generated by \eqref{plagsonadj} and \eqref{plagso8adj}. The generator content and the derivative order matches the counting presented in tables \ref{M-inv-table} and \ref{M-non-inv-table} for $SO(8)$.
\subsubsection{$N=6$}
In this case, we have a quasi non-invariant colour module generated by the scalar Lagrangian 
\be
\tilde{\chi}^{SO(6),(1)}=\ve_{ijklmn}\Phi_1^{ij}\phi_3^{kl}\Phi_2^{m\alpha} \Phi_4^{n\alpha}|_{(+--)}=\ve_{ijklmn}\Phi_1^{ij}\Phi_3^{kl}\Phi_2^{m\alpha} \Phi_4^{n\alpha}+\ve_{ijklmn}\Phi_2^{ij}\Phi_4^{kl}\Phi_1^{m\alpha} \Phi_3^{n\alpha}.
\ee 
This transforms in a ${\bf 3_A}$ of $S_3$. Along with the quasi non-invariant photon structure \eqref{photon-lag2}, this contributes to $\CM^{\rm non-inv}$. 
\bea
{\tilde E}^{(1)}&\simeq& -6F^{ab}_{1}  \partial_a F_{2}^{\mu\nu} \partial_b F_{3}^{\nu\rho} F_{4}^{\rho \mu}|_{(+--)}\\
&=& 6\left(-F^{ab}_{1}  \partial_a F_{2}^{\mu\nu} \partial_b F_{3}^{\nu\rho} F_{4}^{\rho \mu} -F^{ab}_{2}  \partial_a F_{1}^{\mu\nu} \partial_b F_{4}^{\nu\rho} F_{3}^{\rho \mu}+F^{ab}_{3}  \partial_a F_{4}^{\mu\nu} \partial_b F_{1}^{\nu\rho} F_{2}^{\rho \mu}+F^{ab}_{4}  \partial_a F_{3}^{\mu\nu} \partial_b F_{2}^{\nu\rho} F_{1}^{\rho \mu}\right). \nonumber
\eea 
The tensor product of the quasi non-invariant colour and photon modules therefore generates ${\bf 3_A}$ generators at 6 derivatives. The gluon module is generated by the following Lagrangian
	\begin{eqnarray}\label{plagso6adj}
	&&H^{1}_{SO(6)}\\
	&& =\sum_{m,n}  a_{m,n} \prod_{b=1}^m\prod_{c=1}^n \ve^{\a\b\g\d\rho\s}T^e_{\a\b}T^f_{\g\d}T^g_{\rho a} T^h_{ \sigma a} 
	~\left(\partial_{\m_b} \partial_{\n_c} F^e_{ab} \partial_{a}F^f_{jk} \partial^{\m_b}\partial_{b}F^g_{kl} \partial^{\n_c} F^h_{lj} - \partial_{\m_b} \partial_{\n_c} F^f_{ab} \partial_{a}F^e_{jk} \partial^{\m_b}\partial_{b}F^g_{kl} \partial^{\n_c} F^h_{lj}\right). \nonumber
	\end{eqnarray}
In summary for $SO(6)$, the local modules $\CM^{\rm inv}$  are generated by \eqref{plagsonadj} and the local module $\CM^{\rm non-inv}$ is genrated by \eqref{plagso6adj}. The generator content and the derivative order matches the counting presented in tables \ref{M-inv-table} and \ref{M-non-inv-table} for $SO(6)$.
\subsubsection{$N=4$}

For $SO(4)$, we have an extra quasi invariant colour module (\eqref{qiso4sc}) and quasi non-invariant colour module (\eqref{qniso4sc}) compared to $N \geq 9$. These are generated by 
\bea
 \chi^{SO(4),(1)}=\Phi_1\wedge \Phi_2 {\rm Tr}(\Phi_3\Phi_4)|_{\Z_2 \times \Z_2}&=&\Phi_1\wedge \Phi_2 {\rm Tr}(\Phi_3\Phi_4)+\Phi_2\wedge \Phi_1 {\rm Tr}(\Phi_4\Phi_3)\nonumber\\
&+&\Phi_3\wedge \Phi_4 {\rm Tr}(\Phi_1\Phi_2)+\Phi_4\wedge \Phi_3 {\rm Tr}(\Phi_2\Phi_1).
\eea
\bea
{\tilde \chi}^{SO(4),(1)}=\Phi_1\wedge \Phi_2 {\rm Tr}(\Phi_3\Phi_4)|_{(+--)}&=&\Phi_1\wedge \Phi_2 {\rm Tr}(\Phi_3\Phi_4)+\Phi_2\wedge \Phi_1 {\rm Tr}(\Phi_4\Phi_3)\nonumber\\
&-&\Phi_3\wedge \Phi_4 {\rm Tr}(\Phi_1\Phi_2)-\Phi_4\wedge \Phi_3 {\rm Tr}(\Phi_2\Phi_1).
\eea
Both the quasi invariant and the quasi non-invariant colour module transform in ${\bf 3}$.  The tensor product of the quasi invariant colour and photon modules therefore generates $({\bf 3}\oplus{\bf 3})_{\rm photon}\otimes {\bf 3}_{\rm scalar}=4 \cdot {\bf 3} \oplus 2 \cdot {\bf 3_A}$ generators at 4 derivatives and $({\bf 1_S})_{\rm photon}\otimes {\bf 3}_{\rm scalar}= {\bf 3}$ generators at $6$ derivatives. They are generated by the following local Lagrangians.
	\begingroup
	\allowdisplaybreaks
	\begin{eqnarray}\label{plagso4adj}
	G^{15}_{SO(4)} &=& \sum_{m,n}  a_{m,n} \prod_{b=1}^m\prod_{c=1}^n \ve^{\a\b\g\d} T^e_{\a\b} T^f_{\g\d} {\rm Tr} \left( T^g T^h\right) ~\left(\partial_{\m_b} \partial_{\n_c} F^e_{ij} F^f_{ji} \right)  \left(\partial^{\m_b} F^g_{kl} \partial^{\n_c} F^h_{lk} \right) \nonumber\\
	G^{16}_{SO(4)} &=& \sum_{m,n}  a_{m,n} \prod_{b=1}^m\prod_{c=1}^n \left(\ve^{\a\b\g\d} T^e_{\a\b} T^g_{\g\d} {\rm Tr} \left( T^h T^f\right) +\ve^{\a\b\g\d} T^e_{\a\b} T^h_{\g\d} {\rm Tr} \left( T^f T^g\right)\right)\left(\partial_{\m_b} \partial_{\n_c} F^e_{ij} F^f_{ji} \right)  \left(\partial^{\m_b} F^g_{kl} \partial^{\n_c} F^h_{lk} \right) \nonumber\\
	G^{17}_{SO(4)} &=& \sum_{m,n}  a_{m,n} \prod_{b=1}^m\prod_{c=1}^n \left(\ve^{\a\b\g\d} T^e_{\a\b} T^g_{\g\d} {\rm Tr} \left( T^h T^f\right) -\ve^{\a\b\g\d} T^e_{\a\b} T^h_{\g\d} {\rm Tr} \left( T^f T^g\right)\right)~\left(\partial_{\m_b} \partial_{\n_c} F^e_{ij} F^f_{ji} \right)  \left(\partial^{\m_b} F^g_{kl} \partial^{\n_c} F^h_{lk} \right) \nonumber\\
	G^{18}_{SO(4)} &=& \sum_{m,n}  a_{m,n} \prod_{b=1}^m\prod_{c=1}^n \ve^{\a\b\g\d} T^e_{\a\b} T^f_{\g\d} {\rm Tr} \left( T^g T^h\right) ~\left(\partial_{\m_b} \partial_{\n_c} F^e_{ij} F^f_{ji} \right)  \left(\partial^{\m_b} F^g_{kl} \partial^{\n_c} F^h_{lk} \right) \nonumber\\
	G^{19}_{SO(4)} &=& \sum_{m,n}  a_{m,n} \prod_{b=1}^m\prod_{c=1}^n \left(\ve^{\a\b\g\d} T^e_{\a\b} T^g_{\g\d} {\rm Tr} \left( T^h T^f\right) +\ve^{\a\b\g\d} T^e_{\a\b} T^h_{\g\d} {\rm Tr} \left( T^f T^g\right)\right)\left(\partial_{\m_b} \partial_{\n_c} F^e_{ij} F^f_{ji} \right)  \left(\partial^{\m_b} F^g_{kl} \partial^{\n_c} F^h_{lk} \right) \nonumber\\
	G^{20}_{SO(4)} &=& \sum_{m,n}  a_{m,n} \prod_{b=1}^m\prod_{c=1}^n\left(\ve^{\a\b\g\d} T^e_{\a\b} T^g_{\g\d} {\rm Tr} \left( T^h T^f\right) -\ve^{\a\b\g\d} T^e_{\a\b} T^h_{\g\d} {\rm Tr} \left( T^f T^g\right)\right)~\left(\partial_{\m_b} \partial_{\n_c} F^e_{ij} F^f_{ji} \right)  \left(\partial^{\m_b} F^g_{kl} \partial^{\n_c} F^h_{lk} \right) \nonumber\\
	G^{21}_{SO(4)} &=& \sum_{m,n}  a_{m,n} \prod_{b=1}^m\prod_{c=1}^n \ve^{\a\b\g\d} T^e_{\a\b} T^f_{\g\d} {\rm Tr} \left( T^g T^h\right) \nonumber\\
	&&~~~~~~~~~~~~~~~~~~~\left(\partial_{\m_b} \partial_{\n_c} F^e_{ab} \partial_{a}F^f_{jk} \partial^{\m_b}\partial_{b}F^g_{kl} \partial^{\n_c} F^h_{lj} + \partial_{\m_b} \partial_{\n_c} F^g_{ab} \partial_{a}F^h_{jk} \partial^{\m_b}\partial_{b}F^e_{kl} \partial^{\n_c} F^f_{lj}\right).
	\end{eqnarray}
	\endgroup
where we have explained the notation in detail below \eqref{plagsonadj}. The tensor product of the quasi non-invariant colour and photon modules generates a ${\bf 3}$ at 4 derivatives. It is generated by the local Lagrangian,
	\bea\label{plagso4adj2}
	H^1_{SO(4)} &=& \sum_{m,n}  a_{m,n} \prod_{b=1}^m\prod_{c=1}^n \ve^{\a\b\g\d} T^e_{\a\b} T^f_{\g\d} {\rm Tr} \left( T^g T^h\right) \nonumber\\
	&&~~~~~~~~~~~~~~~~~\left(\partial_{\m_b} \partial_{\n_c} F^e_{ab} \partial_{a}F^f_{jk} \partial^{\m_b}\partial_{b}F^g_{kl} \partial^{\n_c} F^h_{lj} - \partial_{\m_b} \partial_{\n_c} F^g_{ab} \partial_{a}F^h_{jk} \partial^{\m_b}\partial_{b}F^e_{kl} \partial^{\n_c} F^f_{lj}\right).
	\eea 
where we have explained the notation in detail below \eqref{plagsonadj}.

In summary for $SO(4)$, the local modules $\CM^{\rm inv}$  are generated by \eqref{plagsonadj}  and \eqref{plagso4adj}, while the local module $\CM^{\rm non-inv}$ is generated by  \eqref{plagso4adj2}. The generator content and the derivative order matches the counting presented in tables \ref{M-inv-table} and \ref{M-non-inv-table} for $SO(4)$.

\subsection{$SU(N)$}\label{egmsnsu}

\subsubsection{$N=3$}
For $SU(3)$, there are a reduction in the number of generators. Using the identity listed in \eqref{su3traceid}, we can immediately see that the following relations between the $\Z_2 \otimes \Z_2$ symmetric local module coming from the following Lagrangian structures 
\bea
G^1_{SU(3)}+G^2_{SU(3)}-\frac{1}{3}\left( \frac{G^4_{SU(3)}}{4}  + G^5_{SU(3)} \right)|_{\Z_2 \otimes \Z_2}=0 \nonumber\\
G^7_{SU(3)}+G^8_{SU(3)}-\frac{1}{3}\left( \frac{G^{10}_{SU(3)}}{4}  + G^{11}_{SU(3)} \right)|_{\Z_2 \otimes \Z_2}=0.
\eea
This implies we can eliminate the two Lagrangian structures, $G^1_{SU(3)}$ and $G^7_{SU(3)}$, transforming in ${\bf 3}$ of $S_3$. We also note that using the identity we can relate the singlet of the module generator from the Lagrangian $G^{13}_{SU(3)}$ and $G^{14}_{SU(3)}$ . If the module generated by $G^{13}_{SU(3)}$ and $G^{14}_{SU(3)}$ be $J^{(i)}_1$ and $J^{(i)}_2$ respectively,   
\be\label{relationsu3}
\sum_{i=1,2,3} \frac{1}{3}J^{(i)}_1 + J^{(i)}_2 =0 
\ee
In summary, the local module $\CM^{\rm inv}$  for $N=3$ continue to be generated by \eqref{plagsonadj} except $G^1_{SU(3)}$ and $G^7_{SU(3)}$ and with the relations \eqref{relationsu3} between the modules generated by  
$G^{13}_{SU(3)}$ and $G^{14}_{SU(3)}$. $\CM^{\rm non-inv}$ continues to be generated by \eqref{lagmnisun}. The generator content and the derivative order matches the counting presented in tables \ref{M-inv-table} and \ref{M-non-inv-table} for $SU(3)$.
\subsubsection{$N=2$}
For $N=2$, the only quasi invariant colour module that contributes is $\xi_{{\bf 3},1}$.
The resulting local modules are only of the type $\CM^{\rm inv}$ .They are generated by \eqref{plagsonadj} except $G^{4,5,6,10,11,12,14}_{SO(N)}$. The generator content and the derivative order  matches the counting presented in tables \ref{M-inv-table} and \ref{M-non-inv-table} for $SU(2)$.

  


\providecommand{\href}[2]{#2}\begingroup\raggedright\endgroup

\end{document}